\documentclass[a4paper,12pt,twoside]{article}

\usepackage{slashed}
\usepackage{amsmath}
\usepackage{wasysym}
\usepackage{latexsym}
\usepackage{graphicx}
\usepackage{amssymb}
\usepackage{amsmath}
\usepackage{mathrsfs}
\usepackage{feynmp}
\usepackage{fancyhdr}
\usepackage{color}
\usepackage[english]{babel}
\usepackage[latin1]{inputenc}
\usepackage{amsmath}
\usepackage{amsfonts}
\usepackage{amssymb}
\usepackage{enumerate}
\usepackage{amsthm}
\usepackage{array}
\usepackage{epsfig}
\usepackage[small,bf]{caption}
\usepackage{appendix}
\usepackage{dsfont}

\usepackage{feynmp}
\usepackage{subfigure}

\usepackage{ae,aecompl}

\usepackage{hyperref}

\newcommand{\beq}{\begin{equation}}
\newcommand{\eeq}{\end{equation}}

\title{{\bf \Large Correlation functions in the non-relativistic AdS/CFT correspondence}}
\date{}

\author{
Carlos A. Fuertes$^a$\footnote{carlos.fuertes@uam.es} \ and Sergej Moroz$^b$\footnote{s.moroz@thphys.uni-heidelberg.de}\\ \\
\small\sl $^a$ Instituto de F\'{\i}sica Te\'orica  IFT UAM/CSIC, \\[-1.5mm]
\small \sl Facultad de Ciencias C-XVI, C.U. Cantoblanco, E-28049
Madrid, Spain\\
\small \sl $^b$  Institut f\"ur Theoretische Physik, \\[-1.5mm]
\small \sl Philosophenweg 16, D-69120 Heidelberg, Germany \\
}

\begin{document}
\begin{titlepage}
\maketitle
\begin{picture}(0,0)(0,0)
\put(350, 250){IFT UAM/CSIC-09-13} 
\end{picture}

\begin{abstract}
  We study the correlation functions of scalar operators in the theory defined as the holographic dual of the
  Schr\"odinger background with dynamical exponent $z=2$ at zero temperature and zero chemical potential. We offer a
  closed expression of the correlation functions at tree level in terms of Fourier transforms of the corresponding
  n-point functions computed from pure AdS in the lightcone frame.
  At the loop level this mapping does not hold and one has to use the full Schr\"odinger background, after proper
  regularization.  We explicitly compute the 3-point function comparing it with the specific 3-point function of the
  non-relativistic theory of cold atoms at unitarity. We find agreement of both 3-point functions, including the part
  not fixed by the symmetry, up to an overall normalization constant.
 \end{abstract}
\thispagestyle{empty}
\setcounter{page}{0}
\end{titlepage}


\tableofcontents

\section{Introduction}

Recently there have been some new exciting developments along the lines of extending the AdS/CFT correspondence
\cite{AdSCFT} to the non-relativistic regime \cite{Son, McGreevy1}.\footnote{ There have been also other proposals extending the
holographic correspondence to Lifshitz-like fixed points \cite{Kachru} and to the context of aging phenomena
\cite{Minic}.} The progress has been encouraged by advances in the
experimental realization and control of cold atom gases at the unitary point \cite{Bloch} where they are described by a
conformal non-relativistic effective theory \cite{NishidaSon}.

Following the work of Son \cite{Son} and Balasubramanian and McGreevy \cite{McGreevy1}, we have at our disposal some toy
model holographic constructions of Schr\"odinger invariant field theories. These holographic duals realize the
Schr\"odinger symmetry algebra as the isometry group, according to what one would expect from the standard AdS/CFT
dictionary. Only for the case of $d=2$-spatial dimensions we have a clear control of the dual QFT, the DLCQ$_\beta$ of
$\mathcal{N}=4$ $SU(N)$ SYM \cite{McGreevy2, Maldacena, Rangamani} that can be extended to the DLCQ$_\beta$ of
$\mathcal{N}=2$ $Sp(N)$ SYM when higher curvature effects are included \cite{Adams}. These theories are very different
from the non-relativistic field theories used to described the cold atoms that have, for $N$ species of atoms, a global
$Sp(2N)$ symmetry \cite{Sachdev,Radzihovsky} and with no gauge symmetry present. So far we cannot take the holographic
theories to be a realistic dual of fermions at unitarity. The thermodynamic observables, although satisfying the
expectations of a Schr\"odinger invariant theory, do not match the ones of the cold atom gases observed in the
laboratory (c.f.~also \cite{Kovtun}). Nevertheless, the holographic constructions provide the starting arena for a
systematic study of possible universal features of the theories in this universality class, along the lines of the KSS
bound. Incidentally the KSS bound is saturated for the Schr\"odinger backgrounds \cite{McGreevy2, Maldacena, Rangamani}
but higher order curvature corrections violate it \cite{Adams}.

In what follows we will take a phenomenological approach and proceed to the study of the correlation functions in the
Schr\"odinger background proposed in \cite{Son, McGreevy1} with dynamical exponent $z=2$, which we will take as the
definition of our dual non-relativistic CFT at zero temperature and zero chemical potential. Analogously to the case of
a (relativistic) conformal invariant theory, the form of the correlation functions is constrained. In the case of the
Schr\"odinger symmetry the constrains are milder and already in the 3-point function an overall function is left
undetermined. One interesting question is to compare whether the physics that one gets from the holographic correlation
functions match our expectations from the non-relativistic theory describing cold atoms. This study will be our primary
concern in this paper.

First, we will study in general the n-point scalar correlation functions in position space. We will not derive the full
bulk-to-boundary and bulk-to-bulk propagators in position space for the Schr\"odinger spacetime. Nonetheless, we will
exploit the following fact to get the computations done at tree level: fields must have a definite non-relativistic mass
momentum at the boundary. The holographic construction is based on the embedding of the Schr\"odinger group into the
conformal group in one higher dimension, where the one extra dimension is the geometrical realization of the central
extension of the Galilean group by the mass operator as momentum. In this sense fields must have a definite momentum
along this extra coordinate at the boundary so as to correspond to an operator with a well-defined non-relativistic
mass.  This will allow us to map the computation of the n-point correlation functions at tree level to momentum
projections of AdS correlation functions in light-cone coordinates. Then we will employ this trick to compute the 2- and
3-point functions. From a QFT point of view, this mapping at tree level is analogous to the equivalence of planar graphs
in a non-commutative and a commutative QFT.\footnote{We thank Mukund Rangamani for this observation.} At least for $d=2$
one can argue \cite{Maldacena} that the dual QFT of the Schr\"odinger spacetime is related to a non-commutative theory
that arises from $\mathcal{N}=4$ SYM, the dual of AdS, giving ground for this equivalence. The results of the
computation of the 2- and 3-point functions will back our belief that the holographic theory given by the Schr\"odinger
background and the theory of cold fermions at unitarity are closely related.  The function not fixed by the
Schr\"odinger symmetry in the 3-point function will be exactly the same as the function obtained in a computation in the
theory of cold atoms. We will also discuss in detail the physical meaning of this function.

The paper is organized as follows: In Section \ref{sect:Sch} we provide a brief summary of the Schr\"odinger group and
the recently proposed holographic realization of it via AdS/CFT. In Section \ref{sect:n-point} we describe a general
recipe for calculation of the scalar n-point functions at tree level and discuss the problems coming from the loops in
the associated Witten diagrams. In Section \ref{sect:2-point} we perform the simplest test of our prescription by
computing the 2-point function. Section \ref{sec:3point} is devoted to the calculation of the 3-point function in the
holographic theory. We also compute the 3-point function in the theory of cold atoms. We compare both results and
analyze the information contained in the correlation function. Finally, in Section \ref{sect:concl} we conclude and
present some outlook.

\section{The Schr\"odinger group and the holographic construction}
\label{sect:Sch}
\subsection{The Schr\"odinger group}

For the non-relativistic systems in $d$ spatial dimensions the geometry of the Galilei group plays a pivotal role. The
group acts on the spatial coordinates $\vec{x}$ and time $t$ as: \beq \label{intro1} (t,\vec{x})\to g(t,
\vec{x})=(t+\beta, \mathscr{R} \vec{x}+\vec{v} t+ \vec{a})\quad, \eeq where $\beta \in \mathds{R}$; $\vec{v}, \vec{a}\in
\mathds{R^{d}}$ and $\mathscr{R}$ is a rotation matrix in $d$ spatial dimensions. In quantum mechanics the Galilei group
acts on the Hilbert space by projective representations. For example, for a scalar complex field $\phi$ of mass $M$
\begin{equation} \label{intro2} 
\phi(t,\vec{x})\to \exp \left[ -\frac{iM}{2}(\vec{v}^2 t+ 2 \vec{v}\cdot \vec{x}) \right]
\phi(g^{-1}(t,\vec{x}))\quad.
\end{equation}
The presence of the mass-dependent phase factor in the transformation law tells us of a superselection rule forbidding
the superposition of states of different masses. By enlarging the Galilei group through a central extension, the
introduction of a mass operator or particle number, we can make the representations unitary \cite{Weinberg}.

The centrally extended Galilean algebra consists of the following generators: one particle number $N$, one time
translation $H$, $d$ spatial translations $P_{i}$, $\frac{d(d-1)}{2}$ spatial rotations $M_{ij}$ and $d$ Galilean boosts
$K_{i}$. The non-trivial commutators are 
\begin{equation} \label{NRCFT1}
  \begin{split}
    & [M_{ij},M_{kl}]=i(\delta_{ik} M_{jl}-\delta_{jk}M_{il}+\delta_{il} M_{kj}-\delta_{jl} M_{ki}) \quad , \\ 
    & [M_{ij},K_{k}]=i(\delta_{ik} K_{j}-\delta_{jk} \quad , K_{i})\quad, \qquad [M_{ij},P_{k}]=i(\delta_{ik}P_{j}-\delta_{jk} P_{i})\quad, \\
    & [P_i,K_j]=-i\delta_{ij}N\quad, \qquad [H,K_{j}]=-iP_{j}\quad.
  \end{split}
\end{equation}

Galilei invariance requires the Bargmann superselection rule for masses to hold \cite{Hagen}. This rule states that
every term in the Lagrangian of the non-relativistic Galilei-invariant theory must conserve the total mass. Hence, for
the scattering process of the incoming particles $a, b, \dots$ into the final particles $a^{\prime}, b^{\prime}, \dots$
\beq \label{intro4a} a+b+\dots \to a^{\prime}+b^{\prime}+\dots \eeq the total mass must be conserved
\beq \label{intro4b} M_a+M_b+\dots = M_{a^{\prime}}+M_{b^{\prime}}+\dots \eeq In the non-relativistic theories the mass
plays the role of the conserved charge.

It is remarkable that the group of spacetime symmetries of the non-interacting Schr\"odinger/ diffusion equation is
larger than the Galilei group and forms the Schr\"odinger group \cite{Hagen, Niederer}. This is the non-relativistic
counterpart of the conformal group. For the dynamical exponent\footnote{which is consistent with the Galilei symmetry}
$z=2$, which determines the relative scaling of time and space coordinates $[t]=[\vec{x}]^{z}$, there are two additional
generators: the scaling generator $D$ and the special conformal generator $C$. The dilatation symmetry scales the time
and spatial coordinates differently in non-relativistic physics: \beq \label{NRCFT3a} (t,\vec{x}) \to (\alpha^2 t,
\alpha \vec{x}) \qquad \alpha \in \mathds{R}.  \eeq The action of the special conformal transformation on time and
spatial coordinates is given by \cite{Hagen}: \beq \label{NRCFT3b} \ (t,\vec{x}) \to (\frac{t}{1-\gamma t},
\frac{\vec{x}}{1-\gamma t}) \qquad \gamma\in\mathds{R}.  \eeq The additional non-trivial commutators of the
Schr\"odinger algebra are
\begin{equation} \label{NRCFT4}
  \begin{split}
    & [P_i,D]=-iP_i \quad, \quad [P_i,C]=-iK_i \quad, \quad [K_i,D]=iK_i \quad,  \\ 
    & [D,C]=-2iC \quad, \quad [D,H]=2iH \quad, \quad [C,H]=iD \quad.  \\
  \end{split}
\end{equation}
The Hamiltonian $H$, scale generator $D$ and
conformal generator $C$ close a subalgebra $sl(2,R)$ of the full Schr\"odinger algebra.

The important point to state is that besides the free Schr\"odinger theory there are known examples of interacting
theories which respect the Schr\"odinger symmetry at the quantum level. These theories are called non-relativistic
conformal field theories (NRCFT) \cite{NishidaSon}. One of them is believed to be two-component cold fermions at
unitarity \cite{NishidaSon}, which is of our main interest and that we will describe in more detail later in section
\ref{sec:3point}.

In close analogy with relativistic conformal field theories it is possible to introduce primary
operators\footnote{quasiprimary in the language of \cite{Henkel1, Henkel2}} in NRCFT \cite{NishidaSon}. A local primary
operator $\mathscr{O}(t,\mathbf{x})$ has a well defined scaling dimension $\Delta_{\mathscr{O}}$ and a particle number
$N_{\mathscr{O}}$: \beq \label{NRCFT6} [D,\mathscr{O}]=i\Delta_{\mathscr{O}} \mathscr{O}, \qquad
[N,\mathscr{O}]=N_{\mathscr{O}}\mathscr{O}, \eeq where $\mathscr{O}\equiv\mathscr{O}(t=0,\mathbf{x}=0)$. The primary
operator $\mathscr{O}$ also commutes with $K_i$ and $C$: \beq \label{NRCFT7} [K_{i},\mathscr{O}]=0 \qquad
[C,\mathscr{O}]=0.  \eeq The primary operator $\mathscr{O}$ with the scaling dimension $\Delta_{\mathscr{O}}$ and the
particle number $N_{\mathscr{O}}$ defines an irreducible representation of the full Schr\"odinger group. The
representation is constructed by taking spatial and time derivatives of a primary operator $\mathscr{O}$.

The Schr\"odinger symmetry and causality condition put powerful constraints on the functional form of the correlation
functions of the primary operators. In \cite{Henkel1, Henkel2} it was demonstrated that the Schr\"odinger invariance
fixes the form of the 2-point scalar correlation function (after Wick rotation) \beq \label{u3} G_{2}(\bar{x}_1, \bar{x}_2)=\langle
\varphi_{1}(\bar{x}_1)\varphi_{2}^{*}(\bar{x}_2)\rangle= \delta_{\Delta_1, \Delta_2} \delta_{M_1, M_2}
\mathscr{C}_{\varphi}\frac{\theta(t_1-t_2)}{(t_1-t_2)^{\Delta_1}}\exp\left[-\frac{M_1}{2}\frac{(\vec{x}_1-\vec{x}_2)^{2}}{t_1-t_2}
\right], \eeq where $\bar{x}_i=(t_i,\vec{x}_i)$, $M_i$ and $\Delta_{i}$ denote the non-relativistic mass and the scaling
dimension corresponding to the operator $\varphi_{i}$. $\mathscr{C}_{\phi}$ denotes a normalization constant.

It is possible to construct a kinematic Schr\"odinger invariant
\beq \label{u3a}
y=\frac{[(\vec{x}_1-\vec{x}_3)(t_2-t_3)-(\vec{x}_2-\vec{x}_3)(t_1-t_3)]^{2}}{(t_1-t_2)(t_1-t_3)(t_2-t_3)}
\eeq
from three spacetime points $\bar{x}_1$, $\bar{x}_2$ and $\bar{x}_3$.
For this reason, the functional form of the 3-point function of scalar operators
$G_{3}(\bar{x}_1, \bar{x}_2, \bar{x}_3)=\langle \varphi_{1}(\bar{x}_1) \varphi_{2}(\bar{x}_2)\varphi_{3}^{*}(\bar{x}_3)\rangle$ is fixed up to a scaling
function $\Psi(y)$
\begin{eqnarray} \label{u4}
&&G_{3}(\bar{x}_1, \bar{x}_2, \bar{x}_3)=\delta_{M_1+M_2,M_3} \theta(t_1-t_3) \theta(t_2-t_3) (t_1-t_3)^{-\Delta_{13,2}/2}
(t_2-t_3)^{-\Delta_{23,1}/2}(t_1-t_2)^{-\Delta_{12,3}/2} \times  \nonumber \\
&&\times\exp\left[ -\frac{M_1}{2}\frac{(\vec{x}_1-\vec{x}_3)^2}{t_1-t_3} -\frac{M_2}{2}\frac{(\vec{x}_2-\vec{x}_3)^2}{t_2-t_3} \right] \Psi\left(\frac{[(\vec{x}_1-\vec{x}_3)(t_2-t_3)-(\vec{x}_2-\vec{x}_3)(t_1-t_3)]^{2}}{(t_1-t_2)(t_1-t_3)(t_2-t_3)} \right),
\end{eqnarray}
where we introduced $\Delta_{ij,k}=\Delta_i + \Delta_{j}-\Delta_{k}$.  The scaling function $\Psi(y)$ is an arbitrary
differentiable function which may have a parametric dependence on the non-relativistic masses $M_1$ and $M_2$.

\subsection{The holographic construction}

The original approach \cite{Son,McGreevy1} to construct the holographic dual of a non-relativistic Schr\"odinger
invariant field theory is based on the realization of the symmetry group as the group of isometries of a geometry rather
than a pure String Theory construction. Here the crucial ingredient, already known long time ago, is the embedding of
the Schr\"odinger algebra as a subalgebra of the (relativistic) conformal algebra in one higher dimension. In
particular, one can characterize the Schr\"odinger algebra as the subalgebra of the conformal group that commutes with
the light-cone momentum (cf.~for example \cite{Son, Henkel2, Aharony:1997an, Duval}).

More explicitly, the conformal algebra is given by
\begin{equation}
  \begin{split}
    & [\tilde{M}^{\mu\nu}, \tilde{M}^{\alpha\beta}] = i (\eta^{\mu\alpha}\tilde{M}^{\nu\beta} + \eta^{\nu\beta}
    \tilde{M}^{\nu\alpha} - \eta^{\mu \beta} \tilde{M}^{\nu\alpha} - \eta^{\nu\alpha} \tilde{M}^{\mu\beta})  \quad,  \\
    & [\tilde{M}^{\mu\nu}, \tilde{P}^{\alpha}] = i (\eta^{\mu\alpha} \tilde{P}^{\nu} - \eta^{\nu\alpha} \tilde{P}^{\mu})  \quad, 
    \\
    & [\tilde{D}, \tilde{P}^\mu] = -i \tilde{P}^\mu \quad, \qquad [\tilde{D}, \tilde{K}^\mu] = i \tilde{K}^\mu  \quad, \\
    & [\tilde{P}^\mu, \tilde{K}^\nu] = -2i (\eta^{\mu\nu}\tilde{D} + \tilde{M}^{\mu\nu}) \quad, \qquad
      [\tilde{K}^{\rho}, \tilde{M}^{\mu\nu}] = -i(\eta^{\rho\mu} \tilde{K}^\nu - \eta^{\rho\nu} \tilde{K^{\mu}}) \quad,
  \end{split}
\end{equation}
where the Greek indices run from 0 to $d+1$. If we introduce light cone coordinates $x^\pm = (x^0 \pm
x^{d+1})/\sqrt{2}$, the subalgebra that commutes with the light-cone momentum $\tilde{P}^+ = (\tilde{P}^0 +
\tilde{P}^{d+1})/\sqrt{2}$ is given by
\begin{equation}
  \begin{split}
    & H = \tilde{P}^- \quad, \qquad P^i = \tilde{P}^i \quad, \qquad M^{ij} = \tilde{M}^{ij} \quad, \\
    & K^i = \tilde{M}^{i+} \quad, \qquad D = \tilde{D} + \tilde{M}^{+-} \quad, \qquad C = \frac{\tilde{K}^+}{2} \quad,
  \end{split}
\end{equation}
where the latin indices run from 1 to $d$.  One can verify that this fulfills the Schr\"odinger algebra, \eqref{NRCFT1}
\eqref{NRCFT4}, with $N = \tilde{P}^+$. In this construction we realize the mass operator at the level of the QFT as the
introduction of an additional dimension. The momentum of the fields along that direction gives us the corresponding
non-relativistic mass (see \cite{Henkel2} for a related discussion).

Following this approach one starts with the AdS metric, the holographic dual of the conformal group, and deforms it so
that only the isometries fulfilling the Schr\"odinger algebra with $z=2$ survive. The resulting metric is \cite{Son, McGreevy1}
\begin{equation}\label{eq:sch_metric1}
 ds^2 = -\beta^2 \frac{dt^2}{u^4} + \frac{-2 dt d\xi + dx^idx^i + du^2}{u^2}
\end{equation}
with $i = 1, \dots, d$. We will denote the space given by this geometry as Sch$_{d+3}$.  The parameter $\beta$ is a
measure of the deformation from pure AdS in the light-cone frame, that corresponds to $\beta = 0$. This metric is
regular everywhere and can be realized as a solution of the Einstein equations supported by an Abelian Higgs model in
the broken symmetry phase \cite{McGreevy1} or equivalently a massive vector field \cite{Son}. For $d=2$ the model can be
realized in a full String Theory construction \cite{McGreevy2, Rangamani, Maldacena}. For recent mathematical discussion
of the metric (\ref{eq:sch_metric1}) in terms of Bargmann space see \cite{Horvathy}.

An important reason to work with the deformed metric and not AdS in the light-cone frame is that, remarkably, the causal
structure of the Schr\"odinger metric is the one of a non-relativistic theory, i.e.~sections of the spacetime with $u$
fixed share the same future and past causal sets. The spacetime is said to be non-distinguishing \cite{Rangamani}.

The isometries of this metric are given by
\begin{equation} 
\begin{split}
  & N = i \partial_\xi \quad, \qquad H = i \partial_t  \quad, \qquad M^{ij} = -i ( x^i\partial_j - x^j\partial_i)  \quad, \\
  & K^i = i ( x^i\partial_\xi + t\partial_i )  \quad, \qquad D = i( 2\partial_t + x^j\partial_j + u\partial_u)  \quad, \\
  & C = i ( t^2 \partial_t + \frac{{x^j}^2 + u^2}{2} \partial_\xi + tx^j\partial_j + tu\partial_u ) \quad.
\end{split}
\end{equation}
They obey the Schr\"odinger algebra with the given identifications.  It is worth stressing that the isometry of
\eqref{eq:sch_metric1} associated with the light-cone momentum, $N \equiv \tilde{P}^+$, the central charge of the
Schr\"odinger algebra, is identified with $\partial_\xi$.  From the quantum field theory point of view, $\tilde{P}^+$ is
the mass operator, so that the momentum along $\partial_\xi$ at the boundary can be identified with the non-relativistic
mass. Applying the usual holographic dictionary the value of a field at the boundary of the spacetime acts as source of
the QFT operator. Since we want to consider operators with a well defined non-relativistic mass, we have to further
impose that the fields in the Schr\"odinger metric have a well defined momentum along the $\partial_\xi$ direction at
the boundary.

The other issue is that the non-relativistic mass can be interpreted as number operator. As such its values should be
discretized. This forces upon us the compactification along the $\partial_\xi$ direction. However this direction is
null. This implies that we have to be extremely careful with the divergences arising from the zero modes
there. Nevertheless, this problem does not appear at the tree level in the correlation functions and we can naively
proceed to analyze them without the need of any regularization. At the loop level we have to regularize the theory,
i.e.~we have to make the compact circle spacelike. The natural way to do it is with the introduction of a non-zero
chemical potential still at zero temperature through the proposed metric \cite{McGreevy2}
\begin{equation} \label{eq:metricT0mu}
 ds^2 = -\beta^2 \frac{dt^2}{u^4} + \gamma^2 u^2 d\xi^2 + \frac{-2 dt d\xi + dx^idx^i + du^2}{u^2}
\end{equation}
and at the end try to arrange things such that we can safely take the zero chemical potential limit $\mu \propto
\gamma^2 \rightarrow 0$.


\section{Scalar n-point functions  in the Schr\"odinger background}
\label{sect:n-point}

In this section we will set up for the computation of the n-point function of scalar operators from the Schr\"odinger
holographic dual in position space. First, we will analyze the case of the contribution from tree-level Witten diagrams,
whose evaluation we will map to Fourier transformations of tree-level AdS correlation functions in the light-cone
frame. Afterward, the case of loops in the Witten diagrams will be considered.

Since only in Euclidean space the computation of the correlation functions is well defined, we will implicitly work in
the Euclidean, upon the usual Wick rotation $t \rightarrow i t$. The Wick rotated version of the Schr\"odinger equation
corresponds to the diffusion equation if we further flip the sign of the non-relativistic mass. The Lorentzian result is
then easily obtained after the inverse Wick rotation from the Euclidean answer since we are at zero temperature. In our
case the Euclidean Schr\"odinger metric, Sch$_{d+3}^E$, corresponds to
 \begin{equation}\label{eq:sch_metric_eucl}
 ds^2_E = \beta^2 \frac{dt^2}{u^4} + \frac{-2 i dt d\xi + dx^idx^i + du^2}{u^2} \quad.
\end{equation}
It is a complex metric. However, as long as the associated action remains real, as it is the case here, this should cause
no trouble \cite{Rangamani, Brown}. In the Lorentzian we have to ask the fields to behave at the boundary as $\phi(X) =
e^{-iM\xi} \phi(\bar{x})$, whereas in the Euclidean they have to behave as $\phi(X) = e^{iM\xi} \phi(\bar{x})$. This way
one recovers a Schr\"odinger equation or a diffusion equation respectively for the fields at the boundary when we start
from a relativistic action in the light-cone frame.

We want to compute the n-point functions of field theory operators dual to scalar fields living in Sch$_{d+3}^E$.  The
action of a complex scalar field will be generically of the form
\begin{equation}
 S = \int d^dx \, \sqrt{g} \left[ \partial_\mu \phi  \partial^\mu \phi^* + m_0^2 |\phi|^2 
   + \mathcal{L}_I
 \right] \quad.
\end{equation}
From this action, according to the usual holographic dictionary, the building blocks that we need to compute in order to
obtain the n-point functions are the bulk-to-boundary and the bulk-to-bulk propagator. For example, the 3-point
amplitude at tree level corresponding to interaction vertices of the form $\mathcal{L}_I = ( \phi_1 \phi_2 \phi_3
+c.c.)$ will be
\begin{equation}
 G_3 (X,Y,Z) = \int \frac{da}{u^{d+3}} K_{1}(a;X) K_{2}(a;Y)  K_{3}(a;Z) \quad,
\end{equation}
where $ K_{i}(a;X_i)$ is the bulk-to-boundary propagator.\footnote{Notation: $a=(t_a, \vec{a},\xi_a, u_a)$,
  $\bar{a}=(t_a,\vec{a})$,  $A=(t_a,\vec{a}, \xi_a)$.}

However, there is something very particular to Sch$_{d+3}^E$ that we can exploit: the fields have a definite
momentum along $\partial_\xi$ at the boundary, that is
\begin{equation}
 \lim_{u\rightarrow 0}\phi(x) = e^{iM\xi} \phi(\bar{x}) \quad.
\end{equation}

\subsection{Tree level}

Let us consider what this means for the computation of the correlators at tree level. At tree level, by definition, we
have no loops and thus all momenta flowing in each propagator is fixed in terms of the momenta of the external
insertions.  Our boundary conditions for the fields are such that they have a definite momentum along the $\partial_\xi$
direction. This implies that all propagators in the tree-level diagrams must have a definite momentum along
$\partial_\xi$. Thus we can use, instead of the full propagators in position space, the propagators that are projected
to have a definite momentum along $\partial_\xi$. Remarkably, one can show that these projected propagators in the
Schr\"odinger spacetime are the same as the projected propagators coming from AdS in the light-cone frame, after a shift
of the relativistic mass $m^2 = m_0^2 + \beta^2 M^2$. This way we can construct the tree-level amplitudes from the AdS
propagators after a momentum projection along $\partial_\xi$. In general we can further state that all we need to do is
to have the corresponding AdS amplitude at tree level in light-cone coordinates and project each operator insertion to
have a momentum along $\partial_\xi$ corresponding to the non-relativistic mass, $M_i$, i.e.

\begin{equation} \label{eq:general_rel}
 \langle \phi_1(\bar{x}_1) \dots \phi_n(\bar{x}_n) \rangle_{\text{Sch}}^{(\text{tree level})} = \int \prod_{k=1}^n d\xi_k\, e^{-i\sum_{j=1}^n ( M_j\xi_j )} \langle
 \phi_1(\bar{x}_1, \xi_1) \dots \phi_n(\bar{x}_n, \xi_n)  \rangle_{\text{AdS}}^{(\text{tree level})}
\end{equation}
The expression is valid for both a compact or non-compact $\partial_\xi$-direction with the evident
modifications.\footnote{When $\xi$ is compact, the integration domain is restricted to the period and one also has to
  impose that the relativistic n-point function be periodic in $\xi_k$.}

\begin{figure}[t]
\begin{center}
\includegraphics[width=2.in]{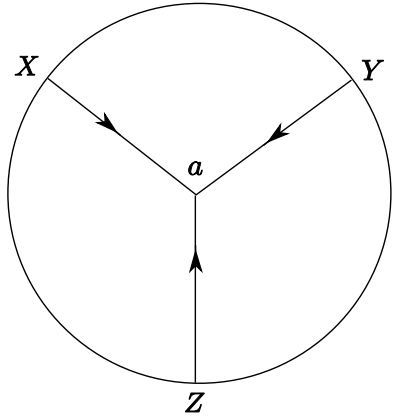}
\end{center}
\vskip -0.7cm \caption{Witten diagram that gives the tree level contribution to the 3-point function.}
\label{fig:3vertex.diag}
\end{figure}

For clarity, let us expound the preceding comments with an example, the 3-point function with contact interactions at
tree level.  In this case, cf.~Fig.~\ref{fig:3vertex.diag}, we will have that the amplitude is obtained
from the functional derivative with respect to $\phi_i(\bar{x}_i)$ of the following integral
\begin{equation}
\begin{split}
 & \int dX \, dY \, dZ  \int  \sqrt{g}\, da\, K_1(a; X) K_2(a; Y) K_3(a; Z) \phi(X) \phi(Y) \phi(Z) = \\
 & = \int d\bar{x} \, d\bar{y} \, d\bar{z}\, d\xi_x \,d\xi_y \,d\xi_z \int \sqrt{g}\, d\bar{a}\,  d\xi_a \, d u_a\, K_1(\bar{a} -
 \bar{x},\xi_a - \xi_x, u_a) K_2(\bar{a} - \bar{y}, \xi_a - \xi_y, u_a ) \times \\ 
   &\qquad \times K_3(\bar{a} - \bar{z}, \xi_a - \xi_z, u_a) e^{iM_1 \xi_x}  e^{iM_2 \xi_y}  e^{iM_{3} \xi_z} \phi(\bar{x}) \phi(\bar{y}) \phi(\bar{z})
\end{split}
\end{equation}
After the functional derivative with respect to $\phi_i(\bar{x}_i)$, a simple change of variables in $\xi_i$ and
integration over $\xi_a$, we obtain that the amplitude is given by
\begin{equation} \label{eq:3-point-amplitude}
\begin{split}
 G_3(\bar{x}, \bar{y}, \bar{z}) =  2\pi\delta(\sum_{k=1}^3 M_k) \int & d\tilde{\xi}_x \,d\tilde{\xi}_y \,d\tilde{\xi}_z
 \int \sqrt{g}\, d\bar{a}\, du_a \, K_1(\bar{a} -  \bar{x},\tilde{\xi}_x, u_a) \times \\ & \times K_2(\bar{a} - \bar{y}, \tilde{\xi}_y, u_a
 ) K_3(\bar{a} - \bar{z}, \tilde{\xi}_z, u_a) e^{-iM_1 \tilde{\xi}_x}  e^{-iM_2 \tilde{\xi}_y}  e^{-iM_{3}\tilde{\xi}_z}
\end{split}
\end{equation}
For definiteness we assumed the $\partial_\xi$-direction to be non-compact as we will do in the rest of
  the expressions in this section when integrating over $\xi$. The same results hold for a compact $\xi$ changing the
  Dirac deltas by Kronecker deltas as well as the overall factor of $2\pi$ by the period along $\partial_\xi$. Also the
  masses then become discrete, $M_i = 2\pi j / L_\xi$ with $j \in \mathds{Z}^*$ and $L_\xi$ the period along
  $\partial_\xi$.

From this example, we see that the Bargmann superselection rule is automatically implemented as the conservation of the
$\partial_\xi$-component of the momentum.  One can associate an arrow with each propagator, such that the ingoing mass
in a vertex should equal the outgoing mass. $\phi$ contributes with $+M$, whereas $\phi^*$ contributes with $-M$. We
also see that the bulk-to-boundary propagator is always going to enter in through the combination
\begin{equation}
\begin{split}
 \phi(\bar{x},\xi, u_x) & = \int d\bar{y} d\xi' K(\bar{x} - \bar{y}, \xi - \xi', u_x) \phi(\bar{y}, \xi')  \\
                   & = e^{i M \xi} \int d\bar{y} d\tilde{\xi} \, K(\bar{x} - \bar{y}, \tilde{\xi}, u_x) \, e^{-iM\tilde{\xi}}
                   \, \phi(\bar{y}) 
\end{split}
\end{equation}
Hence we can introduce what we will call the \emph{projected} bulk-to-boundary propagator
\begin{equation} \label{prbtb}
  K_M (\bar{x}- \bar{y},\xi - \xi', u_x)  =  {e^{iM (\xi - \xi')}} \int d\tilde{\xi} \,  K(\bar{x}-\bar{y},
  \tilde{\xi}, u_x) \, e^{-iM\tilde{\xi}}
\end{equation}
and use it to construct directly the amplitudes in Sch$_{d+3}^E$ from the corresponding Witten diagrams.

Now we can ask, what is the differential equation that the projected bulk-to-boundary propagator satisfies? The
bulk-to-boundary propagator satisfies that
\begin{equation}
 \left( \nabla^2_x - m_0^2 \right) K(x-y) = 0 \quad, \qquad \lim_{u\rightarrow 0} K(x-y) \sim u^{d+2-\Delta} \delta(\bar{x} - \bar{y}) \quad.
\end{equation}

In Sch$_{d+3}^E$ the operator $\left( \nabla^2_x - m_0^2 \right)$ is given by
\begin{equation} \label{sfe}
 \nabla^2_x - m_0^2 =  u^2 \partial_u^2 - (d+1) u \partial_u  + 2i u^2 \partial_t \partial_\xi  + u^2 \partial_i^2 +
 \beta^2 \partial_\xi^2 - m_0^2
\end{equation}
If we consider the projected bulk-to-boundary propagator, we see that it will satisfy
\begin{equation}  \label{eq:op_proj_prop_Sch}
\left[ \partial_u^2  - \frac{d+1}{u} \partial_u - 2M \partial_t + \partial_i^2 -
 \frac{1}{u^2}\left( \beta^2 M^2 + m_0^2  \right) \right] K_M(\bar{x}-\bar{y}, \xi - \xi', u) = 0
\end{equation}

Remarkably, one can compare it to the action of $\left( \nabla^2_x - m^2 \right)$ on the projected propagator but in a
pure AdS$_{p+2}$ background in light-cone coordinates (basically eq.~\eqref{eq:sch_metric_eucl} but with $\beta = 0$),
\begin{equation} \label{eq:op_proj_prop_AdS}
  \left[ \partial_u^2 - \frac{p}{u} \partial_u - 2 M \partial_t  + \partial_i^2  - \frac{m^2}{u^2} \right]
  K_M(\bar{x}-\bar{y}, \xi - \xi',u) = 0
\end{equation}

We see both equations \eqref{eq:op_proj_prop_Sch} and \eqref{eq:op_proj_prop_AdS} are the same provided we identify $p =
d+1$ and take the mass in AdS to be $m^2 = m_0^2 + \beta^2 M^2$. Hence we see that the projected propagator in Sch$_{d+3}$
is the same as the projection of the propagator in AdS$_{d+3}$ in light-cone coordinates with a suitable shift of the
mass,
\begin{equation} \label{eq:propagator}
\begin{split}
 K_M (\bar{x}- \bar{y},\xi - \xi', u_x) = {e^{iM (\xi - \xi')}} \int d\tilde{\xi}\,
 K_\Delta^{(\text{AdS})}(\bar{x}-\bar{y}, \tilde{\xi}, u_x) \, e^{-iM\tilde{\xi}}
\end{split}
\end{equation}
such that 
\begin{equation} \label{ident}
 \Delta(\Delta - d - 2) = m^2 = m_0^2 + \beta^2 M^2 \quad.
\end{equation}

This is nothing but a reflection of the fact that the equations of motion of a free scalar with a definite momentum along
$\partial_\xi$ in Sch$_{d+3}$ are the same as another free scalar in AdS$_{d+3}$ in light-cone coordinates with a
shifted mass and the same momentum along $\partial_\xi$.

Now that we have an integral expression (see Appendix \ref{app:A} for the explicit form) of the projected bulk-to-boundary
propagator in terms of known functions, we can revisit the expression for the 3-point function
\eqref{eq:3-point-amplitude} in terms of the projected bulk-to-boundary propagators, corresponding to the diagram in
Fig.~\ref{fig:3vertex.diag},
\begin{equation} \label{eq:3-point-amplitude2}
\begin{split}
  \langle \phi_1(\bar{x}) & \phi_2(\bar{y}) \phi_3(\bar{z}) \rangle^{(\text{tree level})}_{\text{Sch}}= \\ & = 2\pi
  \delta(\sum_{k=1}^3 M_k) \int \sqrt{g}\, d\bar{a} \,du_a \, K_{M_1}(\bar{a} - \bar{x}, u_a)
  K_{M_2}(\bar{a} - \bar{y}, u_a)  K_{M_3}(\bar{a} - \bar{z}, u_a) \\
  & = 2\pi \delta(\sum_{k=1}^3 M_k) \int d\tilde{\xi_x}\, d\tilde{\xi_y}\, d\tilde{\xi_z} \int \sqrt{g} \, d\bar{a}
  \,du_a \, K_{\Delta_1}^{(\text{AdS})}(\bar{a}-\bar{x}, \tilde{\xi_x}, u_a) \times \\ & \qquad \qquad \qquad \qquad
  \times K_{\Delta_2}^{(\text{AdS})}(\bar{a}-\bar{y}, \tilde{\xi_x}, u_a) K_{\Delta_3}^{(\text{AdS})}(\bar{a}-\bar{z}, \tilde{\xi_z}, u_a) \, e^{-iM\tilde{\xi_x}} e^{-iM\tilde{\xi_y}}  e^{-iM\tilde{\xi_z}} \\
  & = \int d\xi_x\, d\xi_y\, d\xi_z \, e^{-iM_1\xi_x} e^{-iM_2\xi_y} e^{-iM_3\xi_z} \langle \phi_1(\bar{x},\xi_x)
  \phi_2(\bar{y}, \xi_y) \phi_3(\bar{z}, \xi_z) \rangle_{\text{AdS}}^{(\text{tree level})}
\end{split}
\end{equation}
The end result is that we can read off the tree level 3-point function in Sch from the corresponding 3-point function in
AdS space in light-cone coordinates projecting to definite momenta, the non-relativistic masses, along a light-light
direction, as advertised at the beginning.

It is immediate that we can generalize this result to all n-point functions that are built from bulk-to-boundary
propagators, i.e.~only one vertex in the bulk, and to the case of the 2-point function at tree level. The 2-point
function is going to be given as
\begin{equation} \label{eq:2-point}
  \langle \phi_1(\bar{x})  \phi_2(\bar{y}) \rangle^{(\text{tree level})}_{\text{Sch}} =  \int d\xi_x\, d\xi_y \, e^{-iM\xi_x}
e^{-iM\xi_y} \langle \phi_1(\bar{x}, \xi_x)  \phi_2(\bar{y}, \xi_y) \rangle_{\text{AdS}}^{(\text{tree level})}
\end{equation}

\begin{figure}[t]
\begin{center}
\includegraphics[width=2.0in]{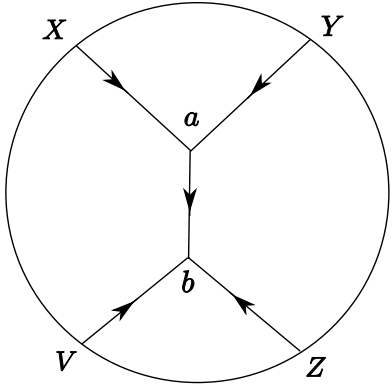}
\end{center}
\vskip -0.7cm \caption{Witten diagram that gives the tree level contribution to the 4-point function with 3-point contact interactions.}
\label{fig:3vertex.4point.diag}
\end{figure}

When the tree-level diagram contains bulk-to-bulk propagators the argument holds the same way.  For example, let us
consider the leading contribution to the 4-point function at tree level from 3-point contact interactions $\mathcal{L}_I
= \phi_1 \phi_2 \phi_3 +c.c.$, corresponding to the diagram in Fig.~\ref{fig:3vertex.4point.diag}.
Denoting by $G(a-b)$ the bulk-to-bulk propagator and again taking into account that $\phi_j(\bar{x}_j, \xi_j) = e^{iM_j \xi_j}
\phi(\bar{x}_j)$ and a simple change of coordinates in $\xi_j$, we have that the amplitude is 
\begin{equation}
\begin{split}
 G_4(\bar{x}, \bar{y}, \bar{z}, \bar{v}) & = - 2\pi \delta(\sum_{j=1}^4 M_j) \int
 g \,d\bar{a}d\bar{b}du_adu_b\,d\tilde{\xi}  K_{M_1}(\bar{a}-\bar{x},\tilde{\xi}_x, u_a) K_{M_2}(\bar{a}-\bar{y},
 \tilde{\xi}_y,u_a) \times \\ & \qquad \times G(\bar{a}-\bar{b},\tilde{\xi}, u_a - u_b) 
  K_{M_3}(\bar{b}-\bar{z},\tilde{\xi}_z, u_b) K_{M_4}(\bar{b}-\bar{v},\tilde{\xi}_v, u_b)  e^{-i(M_3+M_4) \tilde{\xi}}
\end{split}
\end{equation}
We can see that the bulk-to-bulk propagator is projected along $\partial_\xi$ with a mass equal to the mass entering the
vertex that it connects, a reflection that the mass is conserved in the bulk vertices.

In general in the Witten diagrams where there are no loops, the bulk-to-bulk propagator is going to appear always
projected by a mass such that the mass is conserved at each vertex.  Again one can associate an arrow with each
propagator, such that the ingoing mass in a vertex should equal the outgoing mass. $\phi$ contributes with $+M$, whereas
$\phi^*$ contributes with $-M$.

Hence it makes sense to introduce the \emph{projected} bulk-to-bulk propagator as
\begin{equation}
  G_M(\bar{x} -  \bar{y}, \xi - \xi', u_x - u_y)  =  {e^{iM (\xi - \xi')}} \int d\tilde{\xi} \,  G(\bar{x}-\bar{y},
  \tilde{\xi}, u_x - u_y) \, e^{-iM\tilde{\xi}}
\end{equation}

Analogously to the case of the projected bulk-to-boundary propagator it can be shown that the projected bulk-to-bulk
propagator in Sch$_{d+3}$ can be related to a projection of the bulk-to-bulk propagator in AdS$_{d+3}$ written in
light-cone coordinates as
\begin{equation} \label{eq:propagatorbulkbulk}
\begin{split}
 G_M(\bar{x} -  \bar{y}, \xi - \xi', u_x - u_y)  =  {e^{iM (\xi - \xi')}} \int d\tilde{\xi} \,  G_\Delta^{(\text{AdS})}(\bar{x}-\bar{y},
  \tilde{\xi}, u_x - u_y) \, e^{-iM\tilde{\xi}}
\end{split}
\end{equation}
with
\begin{equation}
 \Delta(\Delta - d - 2) = m^2 = m_0^2 + \beta^2 M^2 \quad.
\end{equation}

\begin{figure}[t]
\begin{center}
\includegraphics[width=2.0in]{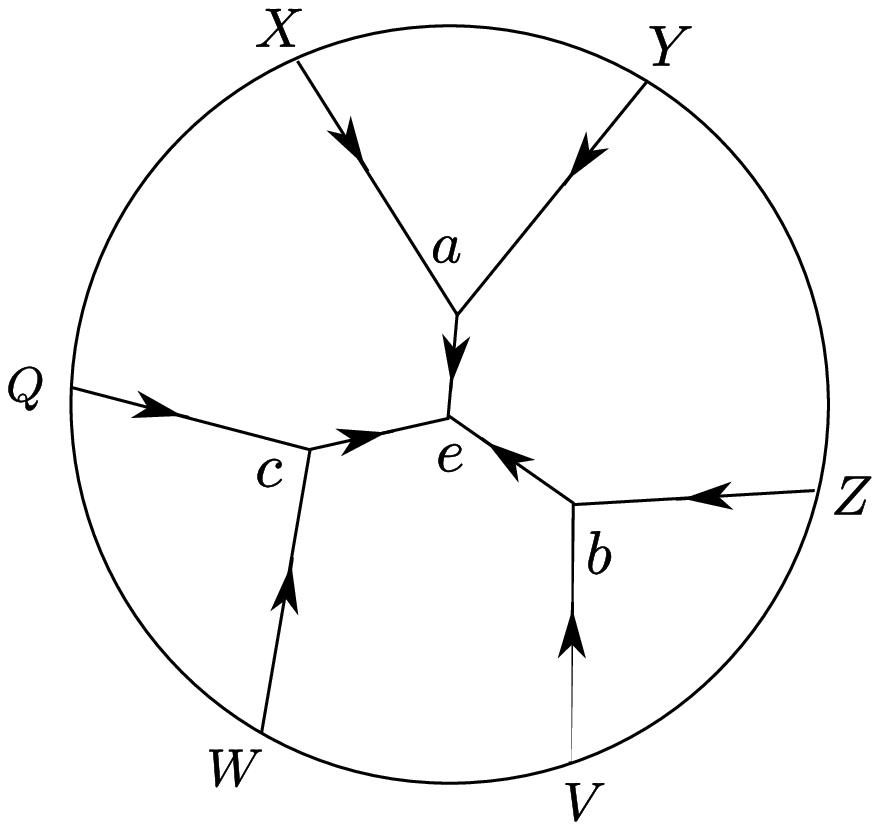}
\end{center}
\vskip -0.7cm \caption{Witten diagram that gives the tree level contribution to the 6-point function with 3-point contact interactions.}
\label{fig:3vertex.6point.diag}
\end{figure}
This way, as another example, the 6-point function at tree level, see Fig.~\ref{fig:3vertex.6point.diag}, is going to be given by
\begin{equation}
\begin{split}
  \langle \phi_1(\bar{x}) & \phi_2(\bar{y}) \phi_3(\bar{z})\phi_4(\bar{v}) \phi_5(\bar{w})\phi_6(\bar{q}) \rangle = \\ &
  = - 2\pi\delta(\sum_{j=1}^6 M_j) \int g^2\, d\bar{a} \,d\bar{b} \,d\bar{c} \,d\bar{e}\, du_a du_b du_c du_e
  K_{M_1}(\bar{a} - \bar{x}, u_a) K_{M_2}(\bar{a} - \bar{y}, u_a) \\ & \quad \times K_{M_3}(\bar{b} - \bar{z}, u_b)
  K_{M_4}(\bar{b} - \bar{v}, u_b) K_{M_5}(\bar{c} - \bar{w}, u_c) K_{M_6}(\bar{c}
  - \bar{q}, u_c) \times \\
  & \quad \times G_{M_1 + M_2}(\bar{e} - \bar{a}, u_e - u_a) G_{M_3 + M_4}(\bar{e} - \bar{b}, u_e - u_b) G_{M_5 + M_6}(\bar{e} - \bar{c}, u_e - u_c)
\end{split}
\end{equation}
where now all functions are known quantities.

Finally we arrive to the observation that in any tree-level diagram we will have that
\begin{equation} \label{eq:Sch_AdS}
 \langle \phi_1(\bar{x}_1) \dots \phi_n(\bar{x}_n) \rangle_{\text{Sch}}^{(\text{tree level})} = \int \prod_{k=1}^n d\xi_k\, e^{-i\sum_{j=1}^n ( M_j\xi_j )} \langle
 \phi_1(\bar{x}_1, \xi_1) \dots \phi_n(\bar{x}_n, \xi_n)  \rangle_{\text{AdS}}^{(\text{tree level})}
\end{equation}
One can easily understand that this must be the case. We take the pure AdS amplitude, built from AdS propagators, but
then we perform non-relativistic mass projections at the boundary that forces all the rest of mass projections in the
internals of the tree level diagram that we see through the projected propagators corresponding to the Sch
amplitude. This way one can understand the pure AdS amplitude as a neat one-higher dimensional representation of the Sch
correlation functions at tree level, which we can read off after appropriate non-relativistic mass projections at the
operator insertions.

What is the QFT viewpoint of relation \eqref{eq:Sch_AdS}? As explained in \cite{Maldacena}, the holographic dual of the
Sch$_{d+3}$ for $d=2$ can be seen as a dipole theory \cite{Alishahiha, Bergman1, Bergman2, Dasgupta} constructed from
$\mathcal{N}=4$ SYM. These dipole theories are a certain type of non-commutative theory that we can construct from an ordinary theory
through the introduction of the non-commutative $*$ product, depending on the parameter $\beta$, given by
\begin{equation}
  f * g = e^{i 2\pi \beta (\tilde{P}^f_- Q^g - \tilde{P}^g_- Q^f) } f g \quad.
\end{equation}
$\tilde{P}_-$ is the light-like momentum charge and $Q$ is another charge associated with some global
symmetries. In the case of $\mathcal{N}=4$ SYM one takes $\tilde{P}_-$ as the charge associated with the momentum along
the light-cone direction $\partial_\xi$ and $Q$ as the charge associated to a $U(1)$ of the $SO(6)$ R-symmetry. The
introduction of this product precisely reduces the spacetime symmetries of the initial SYM theory to the Schr\"odinger
subgroup. The bulk equivalent of the introduction of the non-commutative product consists on the TsT transformation of
the original AdS background that gives the Sch metric along a non-trivial profile for the dilaton and B-field \cite{Maldacena,Bergman1}. In
non-commutative theories the planar-diagrams are simple. They are equal to the planar-diagrams of the ordinary theory
except for some overall phase depending only on the external particles \cite{Filk, Bigatti}. Thus we can
understand that the tree-level contributions (corresponding to the planar diagrams in the QFT) to the correlation
functions are basically the same as computed from Sch or AdS as given by \eqref{eq:Sch_AdS}. 
\smallskip

To summarize we have seen that at tree level, when there are no loops in the associated Witten diagrams, the correlation
function of scalars can be computed in terms of projected propagators coming from pure AdS written in light-cone
coordinates. Furthermore, the amplitude is exactly the mass projection of the equivalent pure AdS amplitude. Hence, at
tree level we can expect no difference in the scalar amplitudes and all derived observables between the results coming
from Sch and light-cone pure AdS provided we shift the mass of the scalar and the fields have a definite momentum along
$\partial_\xi$.

\subsection{Loops}

In the case when there appears loops in the associated Witten diagrams for the amplitudes, we cannot rewrite them in
terms of projected propagators, i.e.~there are undetermined masses over which we must integrate. Hence we cannot compute
loop-amplitudes in the Schr\"odinger spacetime from projections of equivalent full AdS amplitudes. The $1/N$ corrections
of each theory are different. This is in agreement to the non-commutative interpretation of the QFT theory for $d=2$
\cite{Maldacena}, where only at the planar level one has equivalence between the non-commutative and commutative
theories. Furthermore, what becomes evident is that when we choose the $\partial_\xi$ direction to be compact, there
appears divergence associated with the zero-modes of the KK tower on the null $S^1_\xi$, as is common in any DLCQ theory
\cite{Hellerman}. In order to give sense to the theory we must abandon the idea of making the $\partial_\xi$-direction
compact or regularize it through the introduction of a finite chemical potential, \eqref{eq:metricT0mu}, that makes the
$\partial_\xi$-circle spacelike.

\begin{figure}[t]
\begin{center}
\includegraphics[width=2.0in]{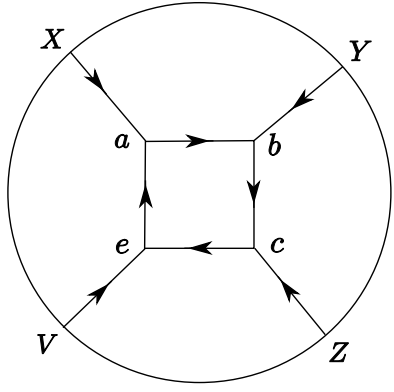}
\end{center}
\vskip -0.7cm \caption{Witten diagram which gives a 1-loop contribution to the 4-point function}
\label{fig:loop.diag}
\end{figure}

As an example let us consider a one loop contribution to the 4-point function. See Fig.~\ref{fig:loop.diag}.
The amplitude is given in terms of
\begin{equation}
\begin{split}
  G_4(\bar{x},\bar{y}, \bar{z}, \bar{v}) = & \int g^2\, d\bar{a}\, d\bar{b}\, d\bar{c}\, d\bar{e}\, d\xi_a d\xi_b d\xi_c
  d\xi_e\, du_a du_b du_c du_e K_{M_1}(\bar{a}-\bar{x}, u_a)  K_{M_2}(\bar{b}-\bar{y}, u_b) \times \\ & \quad \times
  K_{M_3}(\bar{c}-\bar{z}, u_c) K_{M_4}(\bar{e}-\bar{v}, u_e) \mathcal{A} (\bar{a}, \bar{b}, \bar{c}, \bar{e}, u_a, u_b,
  u_c, u_e)
\end{split}
\end{equation}
with
\begin{equation} \label{eq:4loop}
\begin{split}
 \mathcal{A} (\bar{a}, & \bar{b}, \bar{c},  \bar{e},  u_a, u_b, u_c, u_e) = \int d\xi_a d\xi_b d\xi_c d\xi_e\, G(\bar{a}-\bar{b}, \xi_a -
 \xi_b, u_a - u_b ) G(\bar{b}-\bar{c}, \xi_b - \xi_c, u_b- u_c )\times \\ &  \times  G(\bar{c}-\bar{e}, \xi_c - \xi_e, u_c - u_e )  G(\bar{e}-\bar{a},
 \xi_e - \xi_a, u_e - u_a ) e^{-iM_1 \xi_a} e^{-iM_2 \xi_b} e^{-iM_3 \xi_c}
 e^{-iM_4 \xi_e} \\
& = 2\pi \delta(\sum_{j=1}^4 M_j) \int d\tilde{\xi_a}d\tilde{\xi_b} d\tilde{\xi_c}\, G(\bar{a}-\bar{b}, \tilde{\xi}_a,
u_a - u_b )
 G(\bar{b}-\bar{c}, \tilde{\xi}_b, u_b - u_c)  \times \\ & \times G(\bar{c}-\bar{e}, \tilde{\xi}_c, u_c - u_e ) G(\bar{e}-\bar{a},
 -\tilde{\xi}_a -\tilde{\xi}_b -\tilde{\xi}_c, u_e - u_a )
  e^{-iM_1 \tilde{\xi}_a}  e^{-i(M_1+M_2) \tilde{\xi}_b}  e^{-i(M_1 + M_2 + M_3) \tilde{\xi}_c}
\end{split}
\end{equation}
We see that in this case there are bulk-to-bulk propagators that are not projected bulk-to-bulk propagators. This result
can be generalized to any Witten diagram that contains loops. 

If we go to Fourier space we can see clearly how, when the $\partial_\xi$-direction is compact with length $L_\xi$, we
are going to face the usual divergence from the zero modes in the circle, following the same argument as in
\cite{Hellerman}. In momentum-space the bulk-to-boundary and bulk-to-bulk propagators are given by
\begin{equation}
  K^\epsilon(u, k) = \left( \frac{u}{\epsilon} \right)^{\frac{d+2}{2}} \frac{\mathcal{K}_\nu(k u)}{\mathcal{K}_\nu(k \epsilon)} 
\end{equation}
\begin{equation}
  G( u_a, u_b, k) = 
  \left\{ 
    \begin{array}{rcl}
        -(u_a u_b)^{\frac{d+2}{2}} \mathcal{I}_\nu(k u_a) \mathcal{K}_\nu(k u_b) & \mbox{ for} & u_a < u_b \\
         -(u_a u_b)^{\frac{d+2}{2}} \mathcal{K}_\nu(k u_a)  \mathcal{I}_\nu(k u_b)  & \mbox{ for} & u_a > u_b
    \end{array}
   \right.
\end{equation}
such that
\begin{equation} \label{eq:nuk}
  \nu = \left(\frac{(d+2)^2}{4}  + m_0^2 + \beta^2 \frac{n^2}{L^2_\xi} \right)^{\frac{1}{2}} \quad, \qquad k = \sqrt{-2i\frac{n\omega}{L_\xi}+ \vec{p}^2} \quad.
\end{equation}
$\epsilon$ denotes an infinitesimal cutoff at the boundary in the radial coordinate. $n\in \mathds{Z}$ such that $M =
n/L_\xi$ is the mass momentum. $\mathcal{I}_\nu$, $\mathcal{K}_\nu$ are the modified Bessel functions.  The position
space propagators are recovered from
\begin{equation} \label{eq:spacemomentumProps}
\begin{split}
  & K(\bar{a}, \bar{x}, u_a) = (2\pi L_\xi)^{-1}\sum_{n=-\infty}^\infty \int \frac{d^{d} \vec{p}}{(2\pi)^{d}}
  \frac{d\omega}{2\pi}
  e^{-i\vec{p}\cdot(\vec{a} - \vec{x})}  e^{-i\omega(t_a - t_x)} e^{-i \frac{n}{L_\xi} (\xi_a - \xi_x)} K^\epsilon(u_a , k)  \\
  & G(\bar{a}, \bar{b}, u_a , u_b) = (2\pi L_\xi)^{-1}\sum_{n=-\infty}^\infty \int \frac{d^{d} \vec{p}}{(2\pi)^{d}}
  \frac{d\omega}{2\pi} e^{-i\vec{p}\cdot(\vec{a} - \vec{b})} e^{-i\omega(t_a - t_b)} e^{-i \frac{n}{L_\xi} (\xi_a - \xi_b)} G(u_a, u_b,
  k)
\end{split}
\end{equation}
Note that the only difference with respect to the AdS propagators in momentum space \cite{Muck} is the dependence of $\nu$
on the mass momentum $M = n/L_\xi$, as is evident from eq.~\eqref{eq:op_proj_prop_Sch}.

The position propagators can be seen as a sum of propagators with a definite momentum over $S_\xi^1$.  For the
zero-modes, the propagator is going to be proportional to $\delta(t)$, as is immediate from \eqref{eq:nuk},
\eqref{eq:spacemomentumProps}, or even directly from \eqref{eq:op_proj_prop_Sch}. Then a closed loop with two
zero-momentum modes will involve $\delta(t)^2 \propto \delta(0)$, resulting in the typical DLCQ divergence. To
solve it we have to regularize by making the null circle to be space-like. We can do it by turning on a non-zero
chemical potential still at zero temperature by means of the metric \eqref{eq:metricT0mu} introduced in
\cite{McGreevy2}. The other option is to discard altogether the possibility of a compact $\partial_\xi$-direction. The
last option poses the conceptual problem of throwing away a genuine number operator interpretation for the mass operator
of the theory.

\section{Scalar 2-point function}
\label{sect:2-point}

In order to illustrate the general method, introduced in Sec.~\ref{sect:n-point}, we compute first a scalar 2-point
function.  According to Eq.~(\ref{eq:2-point}) the 2-point scalar correlator $G_2=\langle \phi_1(t_1, \vec{x}_1)
\phi_2(t_2, \vec{x}_2) \rangle^{(\text{tree level})}_{\text{Sch}}$ in the non-relativistic holography is given by
\beq \label{2ph1} G_2=\int d\xi_1\, d\xi_2\, e^{-i(M_1\xi_1+M_2\xi_2}) \langle \phi_1(X_1) \phi_2(X_2)
\rangle_{\text{AdS}}^{(\text{tree level})}.  \eeq The relativistic 2-point function in Euclidean space is well-known
from the relativistic AdS/CFT \cite{3-point} \beq \label{2ph2} \langle \phi_1(X_1) \phi_2(X_2)
\rangle_{\text{AdS}}^{(\text{tree level})}=\frac{C_{12}\delta_{\Delta_1, \Delta_2}}{|X_1-X_2|^{2\Delta_{1}}}, \eeq where
$X=(t,\vec{x},\xi )$ and $|X|^{2}=\vec{x}^2-2i t \xi$. $C_{12}$ is a position-independent constant
  which is taken to one under canonical normalization.

After introducing a center of mass coordinate $\eta=\xi_1+\xi_2$ and a relative coordinate $\xi=\xi_1-\xi_2$ we end up with
\beq \label{2ph3}
G_2 =\frac{1}{2}\int d\eta\, d\xi\, e^{-i\frac{M_1+M_2}{2}\eta} e^{-i\frac{M_1-M_2}{2}\xi}  \langle \phi_1(X_1)
\phi_2(X_2)   \rangle_{\text{AdS}}^{(\text{tree level})}.
\eeq
The integration over $\eta$ produces a Bargmann superselection delta function $\delta(M_1+M_2)$. The remaining $\xi$ integral was evaluated in \cite{Henkel2} and can be found in Appendix \ref{app:A}
\begin{eqnarray} \label{2ph4}
G_2&=& 2\pi \delta(M_1+M_2) \int d\xi e^{-iM_1 \xi} \langle \phi_1(X_1)
\phi_2(X_2)   \rangle_{\text{AdS}}^{(\text{tree level})}=  \nonumber \\
&&=\frac{2\pi C_{12}\delta_{\Delta_1, \Delta_2}\delta(M_1+M_2)}{(2i(t_1-t_2))^{\Delta_1}} \int d \xi e^{-iM_1 \xi} \frac{1}{\left(\xi+\frac{i(\vec{x}_1-\vec{x}_2)^2}{2(t_1-t_2)} \right)^{\Delta_1}} = \nonumber \\
&&=\mathscr{C}_{\phi}\delta_{\Delta_1, \Delta_2} \delta(M_1+M_2) \frac{\theta(t_1-t_2)}{(t_1-t_2)^{\Delta_1}} \exp \left(-\frac{M_1}{2}\frac{(\vec{x}_1-\vec{x}_2)^2}{(t_1-t_2)} \right),
\end{eqnarray}
where $\mathscr{C}_{\phi}=\frac{2\pi C_{12}\alpha M_{1}^{\Delta_1-1}}{(2i)^{\Delta}}$. In the case of a
  compact $\partial_\xi$-direction, one arrives at a same expression with the same functional dependence, see Appendix
  \ref{app:compact}. It is reassuring that the final result coincides with Eq.~(\ref{u3}) which follows from the
Schr\"odinger symmetry.


\section{Scalar 3-point function}
\label{sec:3point}

In this Section we tackle a more difficult aim: computation of
the 3-point correlator. First, a specific 3-point function is calculated using QFT methods for cold atoms at
unitarity. Then we employ the general method of Sec.~\ref{sect:n-point} to obtain the scalar 3-point function from
non-relativistic holography. We compare the non-universal scaling functions $\Psi(y)$ (\ref{u4}) for cold atoms and
holographic theory. Finally, we provide a few remarks on physical information stored in the scaling function $\Psi(y)$.

\subsection{3-point function for cold atoms at unitarity}
\label{sec:3pointColdAtoms}

In this subsection we compute the 3-point function in a non-relativistic QFT of cold atoms with contact interaction. The two-component fermions near a broad Feshbach resonance at $T=0$ are described by a microscopic action\footnote{In order to do a direct comparison with \cite{Henkel1, Henkel2} we work in Euclidean time} \cite{Nishida}
\beq \label{u1fa}
S_{E}[\psi]=\int dt \int d^{d} x \sum_{i=1}^{2}\psi_{i}^{*}(\partial_{t}-\frac{\Delta}{2m}-\mu)\psi_{i}-c_0 \psi_{1}^{*} \psi_{2}^{*} \psi_2 \psi_1,
\eeq
where two species of the fermionic atoms of mass $m$ are denoted by $\psi_1$ and $\psi_2$, $\mu$ stands for a chemical potential and $c_0$ characterizes the microscopic interaction strength. The action has an internal $SU(2)\times U(1)$ symmetry. The QFT defined by Eq.~(\ref{u1fa}) must be equipped with a UV cutoff $\Lambda$ due to the contact nature of the interaction term. In this paper we are interested in a vacuum state, i.e. the state of zero temperature and density. In vacuum the bare parameters $\mu$ and $c_0$ are the functions of the cut-off $\Lambda$ and the low-energy s-wave scattering length $a$ only. In order to simplify the analysis, it is useful to rewrite Eq.~(\ref{u1fa}) by means of the Hubbard-Stratonovich transformation:
\beq \label{u1f}
S_{E}[\psi,\phi]=\int dt \int d^{d} x \sum_{i=1}^{2}\psi_{i}^{*}(\partial_{t}-\frac{\Delta}{2m})\psi_{i} +\frac{1}{c_0}\phi^{*}\phi -(\phi^{*}\psi_{1}\psi_{2}+\phi \psi_{2}^{*} \psi_{1}^{*}),
\eeq
where $\phi$ denotes a composite bosonic diatom of mass $2m$. The theory (\ref{u1f}) becomes strongly interacting in the unitary regime $|a|\to\infty$ in $d=3$. In the vacuum state in the unitary limit $\mu=0$ and $c_{0}=c_{0}(\Lambda)$, where the concrete functional form depends on the regularization procedure. For example, for sharp momentum regularization in $d=3$
\beq \label{u1fc}
-\frac{1}{c_0}=\frac{m}{4\pi a}-\int^{\Lambda}\frac{d^{3}q}{(2\pi)^3}\frac{m}{\vec{q}^2}.
\eeq
At unitarity the only scale defined by the scattering length $a$ drops out and the theory becomes classically scale invariant. The QFT defined by Eq.~(\ref{u1f}) is believed to be an example of the non-relativistic CFT which respects the Schr\"odinger spacetime symmetry \cite{NishidaSon}.

The exact Euclidean propagators $G_{\psi}(\omega, \vec{q})$ and $G_{\phi}(\omega, \vec{q})$ can be obtained from (\ref{u1f}) using the non-perturbative Lippmann-Schwinger integral equations. In non-relativistic vacuum there is no particle-antiparticle production and hence the atom propagator $G_{\psi}(\omega, \vec{q})$ is not renormalized. In the momentum space it looks \cite{Nishida}
\beq \label{u1f2a}
G_{\psi}(\omega, \vec{q})=\frac{1}{i \omega+\epsilon_{\vec{q}}} \qquad \qquad \epsilon_{\vec{q}}=\frac{\vec{q}^2}{2m},
\eeq
Quantum effects make the diatom field $\phi$ fully dynamical with the propagator in the scale-free unitary regime given by \cite{Nishida}
\beq \label{u1f2b} 
G_{\phi}(\omega, \vec{q})=\frac{(\frac{4\pi}{m})^{\frac{d}{2}}}{\Gamma(1-\frac{d}{2})} \frac{1}{(i \omega+\frac{\epsilon_{\vec{q}}}{2})^{\frac{d}{2}-1}}.
\eeq
The diatom propagator is strongly renormalized. The form of $G_{\psi}(\omega, \vec{q})$ and $G_{\phi}(\omega, \vec{q})$ is consistent with the Schr\"odinger symmetry leading to the scaling dimensions of the atom field $\psi$ and the diatom field $\phi$
\beq \label{eft5}
\Delta_{\psi}=\frac{d}{2} \qquad \Delta_{\phi}=2.
\eeq
This is in contrast to free fermions where one has that $\Delta_{\psi}=\frac{d}{2}$, $\Delta_{\phi}=d$.

In a very similar fashion one can describe the non-relativistic bosons near a broad Feshbach resonance. The microscopic $U(1)$ symmetric action in vacuum is
\beq \label{u1}
S[\psi,\phi]=\int dt \int d^{d} x \psi^{*}(\partial_{t}-\frac{\Delta}{2m})\psi +\frac{1}{c_0}\phi^{*}\phi+ \frac{1}{2}(\phi^{*}\psi \psi+\phi \psi^{*} \psi^{*}),
\eeq
where $\psi$ represents a complex bosonic atom, while $\phi$ stands for a composite bosonic diatom.
The full Euclidean propagators at unitarity can be calculated exactly in vacuum and are given by
\begin{eqnarray} \label{u2}
G_{\psi}(\omega, \vec{q})&=&\frac{1}{i \omega+\epsilon_{\vec{q}}} \qquad \qquad \epsilon_{\vec{q}}=\frac{\vec{q}^2}{2m} \nonumber \\
G_{\phi}(\omega, \vec{q})&=&\frac{(\frac{4\pi}{m})^{\frac{d}{2}}}{\Gamma(1-\frac{d}{2})} \frac{2}{(i \omega+\frac{\epsilon_{\vec{q}}}{2})^{\frac{d}{2}-1}}.
\end{eqnarray}

While the microscopic actions (\ref{u1f}) and (\ref{u1}) look very similar, the QFTs defined by them are rather different. This is due to the different statistics of atoms in Eqs.~(\ref{u1f}) and (\ref{u1}). Nevertheless, the one-  and two-body sectors\footnote{We define a n-body sector as a set of 2n-point Greens' functions written in terms of elementary atoms. In this sense $\langle \phi^{*}\phi \rangle$ belongs to the two-body sector because $\phi\sim \psi \psi$ is composed of two atoms.} of the bosonic theory (\ref{u1}) have the same form (up to a simple multiplicative factors) as the one- and two-body sectors of the fermionic theory \cite{Sachdev, MFSW} and respect the non-relativistic Schr\"odinger spacetime symmetry. In the three-body sector the bosonic theory (\ref{u1}) exhibits the Efimov effect \cite{Efimov}, i.e.~the non-relativistic conformal anomaly, while there is no Efimov effect for the two-component fermions.

In this subsection we will calculate the specific 3-point function\footnote{ $\bar{x}_i=(t_i,\vec{x}_i)$ in accordance with our notation.} $\langle \psi(\bar{x}_1)\psi(\bar{x}_2)\phi^{*}(\bar{x}_3)
\rangle$   for bosons at unitarity\footnote{This specific 3-point correlator belongs to the two-body sector because it ``consists'' of four elementary fields and hence must be consistent with the non-relativistic conformal form (\ref{u4}). The 3-point correlator $\langle \psi_{1}(\bar{x}_1)\psi_{2}(\bar{x}_2)\phi^{*}(\bar{x}_3) \rangle$ for the fermionic theory (\ref{u1f}) can be calculated along the same lines as the bosonic 3-point correlator \cite{Sachdev, MFSW}. } defined by Eq.~(\ref{u1}), demonstrating the consistency with the non-relativistic conformal form (\ref{u4}) and determining the scaling function $\Psi(y)$. The calculation of the 3-point function can be conveniently done in the position space, hence first we rewrite the propagators (\ref{u2}) in the position representation (for details see Appendix \ref{Fourier})
\begin{eqnarray} \label{u5}
G_{\psi}(t, \vec{x})&=&\int \frac{d \omega}{2\pi} \frac{d^{d}q}{(2\pi)^d} \frac{1}{i \omega+\epsilon_{\vec{q}}} e^{i(\omega t - \vec{q} \cdot \vec{x})}
= C_{\psi}\theta(t) t^{-\frac{d}{2}}\exp \left( -\frac{m}{2} \frac{\vec{x}^2}{t} \right), \nonumber \\
G_{\phi}(t, \vec{x})&=&\frac{(\frac{4\pi}{m})^{\frac{d}{2}}}{\Gamma(1-\frac{d}{2})}\int \frac{d \omega}{2\pi} \frac{d^{d}q}{(2\pi)^d} \frac{2}{(i \omega+\frac{\epsilon_{\vec{q}}}{2})^{\frac{d}{2}-1}} e^{i(\omega t - \vec{q} \cdot \vec{x})}=C_{\phi}\theta(t) t^{-2}\exp \left( -m \frac{\vec{x}^2}{t} \right),
\end{eqnarray}
where $C_{\psi}=\left(\frac{m}{2\pi} \right)^{\frac{d}{2}}$ and $C_{\phi}=(4)^{\frac{d}{2}}\frac{\left(d-2 \right)\sin
  \left(\frac{\pi}{2}d \right)}{\pi}$. The last equation is in agreement with Eq.~(\ref{u3}) after the
identification $M_{\psi}=m$ and $M_{\phi}=2m$. The unitarity scaling dimensions of $\psi$ and $\phi$ can be easily read
off from Eq.~(\ref{u5}): $\Delta_{\psi}=\frac{d}{2}$ and $\Delta_{\phi}=2$.

\begin{figure}[t]
\begin{center}
\includegraphics[width=2.0in]{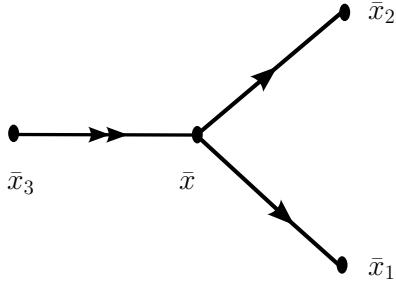}
\end{center}
\vskip -0.7cm \caption{A Feynman diagram which gives the 3-point function for atoms at unitarity. The lines with a single arrow denote the exact atom propagator $G_{\psi}$, while the line with two arrows represents the fully renormalized diatom propagator $G_{\phi}$.}
\label{fig1}
\end{figure}

Now we are ready to calculate the 3-point function $\langle \psi(\bar{x}_1)\psi(\bar{x}_2)\phi^{*}(\bar{x}_3) \rangle$ which corresponds to scattering of two atoms $\psi$ into a diatom $\phi$. The non-relativistic Bargmann superselection rule for masses is satisfied $m+m-2m=0$ and hence the 3-point function is non-trivial. There are two important observations which allow us to calculate the full 3-point function
\begin{itemize}
\item There is no condensate in vacuum $\langle \psi \rangle=0$, $\langle \phi \rangle=0$. This implies that the full 3-point function is given by the connected part only.
\item The Yukawa vertex is not renormalized in the non-relativistic vacuum \cite{Sachdev, MFSW}. Hence only one Feynman diagram contributes to the full 3-point function (Fig.~\ref{fig1}). The specific 3-point correlator is completely determined by the full Euclidean propagators (\ref{u2}) of atoms and diatoms, i.e.~by the scaling dimensions of the fields.
\end{itemize}
From translation invariance the 3-point function depends only on the distances $\bar{x}_1-\bar{x}_3$, $\bar{x}_2-\bar{x}_3$ and $\bar{x}_2-\bar{x}_1$. Setting $\bar{x}_3=(t_3,\vec{x}_3)=0$ we obtain
\begin{equation} \label{u6}
\begin{split}
G_3(\bar{x}_1,\bar{x}_2)&=\int d\bar{x} G_{\phi}(\bar{x}) G_{\psi}(\bar{x}_1-\bar{x}) G_{\psi}(\bar{x}_2-\bar{x})=\\
&=C_{\psi}^{2}C_{\phi}\int d^{d} x \int dt \theta(t) \theta(t_1-t) \theta(t_2-t) t^{-2} (t_1-t)^{-\frac{d}{2}}
(t_2-t)^{-\frac{d}{2}} \times  \\
& \quad \quad \quad \quad \times \exp \left[-\frac{m}{2} \left( \frac{2\vec{x}^2}{t}+\frac{(\vec{x}_1-\vec{x})^{2}}{(t_1-t)}+ \frac{(\vec{x}_2-\vec{x})^{2}}{(t_2-t)} \right) \right].
\end{split}
\end{equation}
The product of three $\theta$ functions in Eq.~(\ref{u6}) is non-vanishing only if $t_{1}>0$ and $t_2>0$. This is in agreement with the causal factor $\theta(t_1-t_3) \theta(t_2-t_3)$ in Eq.~(\ref{u4}). The product $\theta(t) \theta(t_1-t) \theta(t_2-t)$ in Eq.~(\ref{u6}) restricts the time integration domain to $t\in (0, \text{min}(t_1,t_2))$. Without loss of generality we take $t_1>t_2$ yielding $t\in(0, t_2)$ in Eq.~(\ref{u6}).

We first perform the Gaussian spatial integration in $d$ dimensions
\begin{eqnarray} \label{u7}
G_{3}(\bar{x}_1, \bar{x}_2)=\left(\frac{2\pi}{m} \right)^{\frac{d}{2}} C_{\psi}^{2} C_{\phi} \theta(t_1) \theta(t_2)
\int_{0}^{t_2} dt \tau^{\frac{d}{2}} t^{-2} (t_1-t)^{-\frac{d}{2}} (t_2-t)^{-\frac{d}{2}} \times \nonumber \\ 
\times \exp \left[\frac{m}{2}\left( \tau \left[ \frac{\vec{x}_1}{t_1-t}+ \frac{\vec{x}_2}{t_2-t} \right]^2 -\frac{\vec{x}_{1}^{2}}{t_1-t}-\frac{\vec{x}_{2}^{2}}{t_2-t} \right) \right],
\end{eqnarray}
where we introduced $\tau=\left( \frac{2}{t}+\frac{1}{t_1-t}+\frac{1}{t_2-t} \right)^{-1}=\frac{t(t_1-t)(t_2-t)}{2t_1 t_2 -t (t_1+t_2)}$.

At this point it is convenient to perform a substitution $t\to z=\frac{t(t_2-t_1)}{t(t_1+t_2)-2t_1 t_2}$. The new dimensionless variable $z$ ranges in the interval $z\in(0,1)$. The argument of the exponential in (\ref{u7}) can now be conveniently rewritten as
\beq \label{u8}
-\frac{m}{2}\frac{\vec{x}_{1}^{2}}{t_1}-\frac{m}{2}\frac{\vec{x}_{2}^{2}}{t_2}-\underbrace{\frac{m}{2}\frac{(\vec{x}_1 t_2- \vec{x}_2 t_1)^2}{t_1 t_2 (t_1-t_2)}}_{y} z.
\eeq
Due to our assumption $t_1>t_2$, a new variable $y$ is non-negative $y\ge 0$.

The final integration can now be done straightforwardly with the result
\begin{equation}  \label{u9}
G_{3}(\bar{x}_1, \bar{x}_2)= \frac{\theta(t_1) \theta(t_2)}{t_1 t_2 (t_1 -t_2)^{\frac{d}{2}-1}}
\exp \left( -\frac{m}{2}\frac{\vec{x}_{1}^{2}}{t_1} -\frac{m}{2}\frac{\vec{x}_{2}^{2}}{t_2} \right) \underbrace{ \left( \frac{2\pi}{m} \right)^{\frac{d}{2}} \frac{C_{\psi}^2 C_{\phi}}{2}\int_{0}^{1} dz z^{\frac{d}{2}-2}\exp(-y z)}_{\Psi(y)}.
\end{equation}
After recalling that in our case $\Delta_{12,3}=d-2$ and $\Delta_{13,2}=\Delta_{23,1}=2$, we observe that the final formula agrees with the Henkel's prediction (\ref{u4}) after recovering $\bar{x}_3$ coordinates: $\bar{x}_1\to \bar{x}_1-\bar{x}_3$ and $\bar{x}_2\to \bar{x}_2-\bar{x}_3$.

We are now in position to determine the  non-universal scaling function $\Psi(y)$ for $y\ge 0$
\begin{equation} \label{u10}
\begin{split}
\Psi(y)&=\left( \frac{2\pi}{m} \right)^{\frac{d}{2}} \frac{C_{\psi}^2 C_{\phi}}{2}\int_{0}^{1} dz z^{\frac{d}{2}-2}\exp (-y z)=  \\
&=\left( \frac{2\pi}{m} \right)^{\frac{d}{2}} \frac{C_{\psi}^2 C_{\phi}}{2} y^{-\frac{d}{2}+1} \gamma(\frac{d}{2}-1,y)\quad,
\end{split}
\end{equation}
where the second line is valid for $d>2$ and a lower incomplete gamma function $\gamma(n,y)$ is defined by
\beq \label{u11}
\gamma(n,y)=\int_{0}^{y}t^{n-1}e^{-t}dt\quad.
\eeq 

We remark that it is possible to generalize the theory (\ref{u1fa}) to a model with $N$ fermion flavors \cite{Sachdev, Radzihovsky}:
\beq \label{u12}
S_{E}^{N}[\psi]=\int dt \int d^{d} x \sum_{\alpha=1}^{N} \sum_{i=1}^{2}\psi_{i\alpha}^{*}(\partial_{t}-\frac{\Delta}{2m}-\mu)\psi_{i\alpha}-\frac{c_0}{N}\sum_{\alpha,\beta=1}^{N} \psi_{1\alpha}^{*} \psi_{2\alpha}^{*} \psi_{2\beta} \psi_{1\beta}.
\eeq
For $N=1$ one recovers the original (\ref{u1fa}) theory. The theory is invariant under $U(1)\times Sp(2N)$ internal group and admits a sensible $1/N$ expansion. The calculation of the 3-point function $G_{3}^{N}=\langle \psi_{1\alpha}(\bar{x}_1)\psi_{2\alpha}(\bar{x}_2)\phi^{*}(\bar{x}_3) \rangle$ can be done straightforwardly with the result $G_{3}^{N}=N G_{3}^{N=1}$.

\subsection{3-point function from non-relativistic holography}
According to Eq.~(\ref{eq:3-point-amplitude2}) the 3-point correlator $G_3=\langle \phi_1(t_1, \vec{x}_1)  \phi_2(t_2, \vec{x}_2)   \phi_3(t_3, \vec{x}_3)  \rangle^{(\text{tree level})}_{\text{Sch}}$ in the non-relativistic holography is given by
\beq \label{3ph1}
G_3=\int d\xi_1\, d\xi_2\, d\xi_3 \, e^{-i(M_1\xi_1+M_2\xi_2+M_3\xi_3)}  \langle \phi_1(X_1)
\phi_2(X_2)   \phi_3(X_3)  \rangle_{\text{AdS}}^{(\text{tree level})}.
\eeq
When $\partial_\xi$ is not compact, the conformal 3-point function in Euclidean space is well-known from the relativistic AdS/CFT \cite{3-point, Muck,
3-point2}
\beq \label{3ph2}
\langle \phi_1(X_1)
\phi_2(X_2)   \phi_3(X_3)  \rangle_{\text{AdS}}^{(\text{tree level})}=\frac{C_{123}}{|X_1-X_2|^{\Delta_{12,3}}|X_2-X_3|^{\Delta_{23,1}}|X_3-X_1|^{\Delta_{31,2}}},
\eeq
here $X=(t,\vec{x},\xi )$, $|X|^{2}=\vec{x}^2-2i t \xi$ and $\Delta_{ij,k} = \Delta_i + \Delta_j -
  \Delta_k$. $C_{123}$ is a position-independent coefficient which is a function of the scaling dimensions
$\Delta_i$, given under canonical normalization by
  \begin{equation}
    C_{123} = - \frac{\Gamma(\frac{\Delta_{12,3}}{2}) \Gamma(\frac{\Delta_{13,2}}{2}) \Gamma(\frac{\Delta_{23,1}}{2})
      \Gamma(\frac{\Delta_1 + \Delta_2 + \Delta_3 - 4 }{2})}{2 \pi^4 \Gamma(\Delta_1 - 2) \Gamma(\Delta_2 - 2)
      \Gamma(\Delta_3 - 2) } \quad.
  \end{equation}

It is convenient to introduce a center-of-mass and relative coordinates
\beq \label{3ph3}
\eta=\xi_1+\xi_2+\xi_3 \qquad \xi=\xi_1-\xi_3 \qquad \xi^{\prime}=\xi_2-\xi_3.
\eeq 
The integration over the center-of-mass coordinate $\eta$ produces a $\delta(M_1+M_2+M_3)$ which is a Bargmann superselection rule:
\beq \label{3ph4}
G_3 =2\pi \delta(M_1+M_2+M_3)\int d\xi\, d\xi^{\prime}\, e^{-i(M_1\xi+M_2 \xi^{\prime})}  \langle \phi_1(X_1)
\phi_2(X_2)   \phi_3(X_3)  \rangle_{\text{AdS}}^{(\text{tree level})}.
\eeq
Now Eq.~(\ref{3ph2}) can be substituted into Eq.~(\ref{3ph4}) and remaining integrals in Eq.~(\ref{3ph4}) were evaluated in the Appendix of \cite{Henkel2}
\begin{equation} \label{3ph5}
\begin{split}
G_3 & = \delta(M_1+M_2+M_3) \theta(t_1-t_3) \theta(t_2-t_3) (t_1-t_3)^{-\Delta_{13,2}/2} (t_2-t_3)^{-\Delta_{23,1}/2}(t_1-t_2)^{-\Delta_{12,3}/2} \times  \\
 & \times \exp\left[ -\frac{M_1}{2}\frac{(\vec{x}_1-\vec{x}_3)^2}{t_1-t_3} -\frac{M_2}{2}\frac{(\vec{x}_2-\vec{x}_3)^2}{t_2-t_3} \right] \Psi\left(\frac{[(\vec{x}_1-\vec{x}_3)(t_2-t_3)-(\vec{x}_2-\vec{x}_3)(t_1-t_3)]^{2}}{(t_1-t_2)(t_1-t_3)(t_2-t_3)} \right)
\end{split}
\end{equation}
The scaling function $\Psi(y)$, which is not directly fixed by the Schr\"odinger symmetry, has the integral
representation \cite{Henkel2}: \beq \label{3ph6} \Psi(y)=\tilde{C}_{123}\int_{\mathds{R}+i\epsilon} dv\,
\int_{\mathds{R}+i\epsilon^{\prime}} dv^{\prime}\, e^{-iM_1 v-iM_2 v^{\prime}} (v-v^{\prime}+iy)^{-\Delta_{12,3}/2}
(v^{\prime})^{-\Delta_{23,1}/2} v^{-\Delta_{13,2}/2}, \eeq where $\tilde{C}_{123}=2\pi C_{123}
(-2i)^{-\frac{1}{2}(\Delta_1+\Delta_2+\Delta_3)}$ and the last expression is valid for $y\in
\mathds{R}$. See appendix \ref{app:compact} for the case of a compact $\partial_\xi$-direction.

\subsection{Comparison of the scaling functions}

Now we can compare the non-universal scaling functions $\Psi(y)$ of cold atoms and non-relativistic holography. To do so we first introduce a related function $\Phi(y)$
\begin{equation} \label{comp1}
\begin{split}
G_{3} & =\delta(M_1+M_2+M_3) \theta(t_1-t_3) \theta(t_2-t_3) (t_1-t_3)^{-\Delta_{13,2}/2} (t_2-t_3)^{-\Delta_{23,1}/2}|t_1-t_2|^{-\Delta_{12,3}/2} \times \\
&\times \exp\left[ -\frac{M_1}{2}\frac{(\vec{x}_1-\vec{x}_3)^2}{t_1-t_3} -\frac{M_2}{2}\frac{(\vec{x}_2-\vec{x}_3)^2}{t_2-t_3} \right] \Phi\left(\frac{[(\vec{x}_1-\vec{x}_3)(t_2-t_3)-(\vec{x}_2-\vec{x}_3)(t_1-t_3)]^{2}}{(t_1-t_2)(t_1-t_3)(t_2-t_3)} \right)
\end{split}
\end{equation}
In comparison with (\ref{u4}) we changed $(t_1-t_2)\to|t_1-t_2|$. This implies $\Phi(y)=\Psi(y)$ for $y\ge 0$ and $\Phi(y)=\Phi(-y)$ because the full 3-point function is symmetric under $t_1\leftrightarrow t_2$. In Euclidean QFT the scaling function $\Phi(y)$ must be real for $y\in \mathds{R}$.

\begin{figure}[t]
\begin{center}
\includegraphics[width=3.0in]{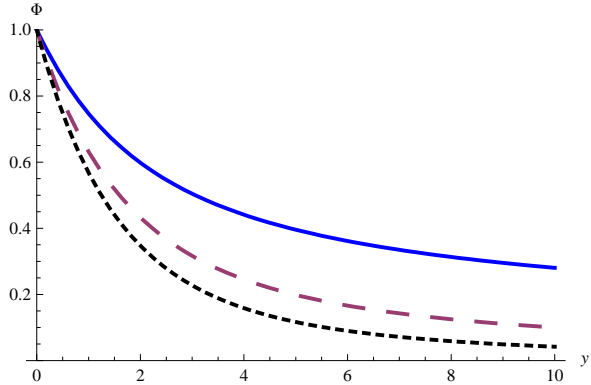}
\end{center}
\vskip -0.7cm \caption{The normalized scaling function $\Phi(y)$ for various dimensions $d=3$ (solid),
  $d=4$ (dashed) and $d=5$ (dotted).}
\label{fig4}
\end{figure}

For cold atoms the analytic expression for the scaling function was found in Eq.~(\ref{u10}). To achieve a simple comparison with the holographic calculation we normalize $\Phi(y)$ such that $\Phi(y=0)=1$. We plot the normalized scaling function for spatial dimensions $d=3, 4, 5$ taking $m=1$ in Fig.~\ref{fig4}.

In the holographic case the scaling function has the integral representation (\ref{3ph6}). To perform a direct
comparison with cold atoms we evaluate Eq.~(\ref{3ph6}) for $\Delta_1=\Delta_2=\frac{d}{2}$, $\Delta_3=2$ and
$M_1=M_2=1$. For general even dimension $d=2n$ the two integrals in Eq.~(\ref{3ph6}) can be done analytically using the residue theorem. Employing a useful relation
\beq \label{comp2}
\gamma(n,y)=(n-1)!+\frac{\partial^{n-1}}{\partial^{n-1}s} \left(\frac{e^{ys}}{s} \right) \Big |_{s=-1} \qquad n\in \mathds{N},
\eeq
which directly follows from the definition (\ref{u11}) we arrive at
\beq \label{comp3}
\Phi(y)=N_{d}y^{-\frac{d}{2}+1}\gamma(\frac{d}{2}-1, y),
\eeq
where $N_d$ is a $y-$independent numerical factor.  This agrees with the cold atoms expression
(\ref{u10}) up to normalization. In
the case of odd dimensions $d=2n+1$ the integral (\ref{3ph6}) has a branch cut and one has to use the integral
representation of the incomplete gamma function \cite{Gradshteyn}
\begin{equation} \label{eq:int_rep_gamma}
  \gamma(\alpha,x) = \Gamma(x) - \frac{e^{-x}x^\alpha}{\Gamma(1-\alpha)} \int_0^{\infty} \frac{e^{-t}
    t^{-\alpha}}{x+t}dt \quad,
\end{equation}
arriving at the same result \eqref{comp3}. Thus for even and odd dimensions the scaling function $\Phi(y)$, calculated
from non-relativistic holography, agrees with $\Phi(y)$ for cold atoms. In the case of a compact
  $\partial_\xi$ one also finds agreement, see Appendix \ref{app:compact}.

At this point it is important to stress that, although the original motivation for the non-relativistic holography were
fermions at unitarity, there are various thermodynamic evidence by now \cite{Kovtun, Chen} that the holography, which
defines some Schr\"odinger invariant field theory, is not exactly dual to the theory of unitary fermions. It is
interesting that our vacuum calculation of the 3-point scaling function $\Phi(y)$, which is not fixed by the
Schr\"odinger symmetry and is thus non-universal, gives the same result for cold atoms and holography. In this respect
it would be interesting to study and compare the higher point functions in both theories.

\subsection{Physical meaning of the scaling function}
The general form of the non-relativistic 3-point function (\ref{comp1}) is quite complex and in order to gain some understanding of the physical information stored in it (and especially in the scaling function  $\Phi(y)$) we must take a specific kinematic configuration of $\bar{x}_1$, $\bar{x}_2$ and $\bar{x}_3$. First, by translational invariance we can take $\bar{x}_3=(t_3, \vec{x}_3)=0$ and the general form (\ref{comp1}) simplifies to
\begin{equation} \label{pm1}
\begin{split}
G_{3}(\bar{x}_1, \bar{x}_2, \bar{x}_3=0) = \, & \delta(M_1+M_2+M_3) \theta(t_1) \theta(t_2) (t_1)^{-\Delta_{13,2}/2}
(t_2)^{-\Delta_{23,1}/2}|t_1-t_2|^{-\Delta_{12,3}/2} \times \\
& \times \exp\left[ -\frac{M_1}{2}\frac{\vec{x}_{1}^{2}}{t_1} -\frac{M_2}{2}\frac{\vec{x}_{2}^2}{t_2} \right] \Phi\left(
  y \right) \quad,
\end{split}
\end{equation}
where the scaling argument $y$ is given by
\beq \label{pm2}
y=\frac{[\vec{x}_1 t_2-\vec{x}_2 t_1]^{2}}{t_1t_2(t_1-t_2)}=\frac{1}{2}\left( \frac{(\vec{x}_1-\vec{x}_2)^2}{t_1-t_2}+\frac{\vec{x}_{2}^{2}}{t_2}-\frac{\vec{x}_{1}^2}{t_1} \right)\quad.
\eeq
In order to simplify things even further we assume that $\vec{x}_1$ and $\vec{x}_2$ lay on the same ray, which starts at the origin\footnote{This choice makes the problem essentially one-dimensional.}, and that $\bar{x}_2$ scales non-relativistically, i.e. $\bar{x}_2=(l^2, \vec{l})$, where $l$ is some length scale assume to be $O(1)$.

In general, the non-relativistic scaling of time and space separation is substantial in the analysis of the structure of the non-relativistic Greens' functions. For example, the generic fall-off of the 2-point function (\ref{2ph4}) as $\bar{x}_1-\bar{x}_2\to\infty$ is exponential. However, if one takes $(\vec{x}_1-\vec{x}_2)^2\sim (t_1-t_2)$, i.e.~applies the non-relativistic scaling, the propagator (\ref{2ph4}) decays by a power-law. 

We choose the point $\bar{x}_1$ to be close to the point $\bar{x}_2$,
i.e. $\bar{x}_1=(l^2(1+\epsilon_t),\vec{l}(1+\epsilon_x))$, where $\epsilon_t, \epsilon_x>0$ and $\epsilon_t, \epsilon_x \ll 1$. Both (\ref{pm1}) and (\ref{pm2}) can be expanded in the small quantities $\epsilon_t, \epsilon_x$
\begin{eqnarray} \label{pm3}
G_3(l,\epsilon_t, \epsilon_x)&\sim& l^{-\sum \Delta_i}\epsilon_{t}^{-\frac{\Delta_{12,3}}{2}}\Phi(y)[1+O(\epsilon_x, \epsilon_t)] \nonumber \\
y&=&\frac{1}{2}\frac{\epsilon_{x}^{2}}{\epsilon_t}+O(\epsilon_x, \epsilon_t).
\end{eqnarray}
From the last expression it is clear that the scaling function $\Phi(y)$ determines the singular behavior of the 3-point function $G_3$ as $\bar{x}_1\to \bar{x}_2$.

In order to illustrate this fact consider our special case with $\Phi(y)$ given by (\ref{comp3}), $M_1=M_2=1$ and $\Delta_{12,3}=d-2$
\beq \label{pm4}
G_3(l,\epsilon_t,\epsilon_x)\sim l^{-d-2} \epsilon_{x}^{-d+2}\gamma(\frac{d}{2}-1, \frac{1}{2}\frac{\epsilon_{x}^2}{\epsilon_t}).
\eeq
We can distinguish two limits:
\begin{itemize}
\item $\epsilon_t\gg \epsilon_x$, i.e. $\bar{x}_1$ approaches $\bar{x}_2$ much faster in the spatial than in the temporal direction. Using the limit
\beq \label{pm5}
\lim_{y\to 0}\gamma(a,y)= a^{-1}y^{a}
\eeq
we arrive at
\beq \label{pm6}
G_3(l,\epsilon_t,\epsilon_x)\sim l^{-d-2} \epsilon_{t}^{1-\frac{d}{2}}.
\eeq
\item $\epsilon_t\ll \epsilon_x$, i.e. $\bar{x}_1$ approaches $\bar{x}_2$ much faster in the temporal than in the spatial direction. In this case
\beq \label{pm7}
\lim_{y\to \infty}\gamma(a,y)= \Gamma(a)
\eeq
and we end up with
\beq \label{pm8}
G_3(l,\epsilon_t,\epsilon_x)\sim l^{-d-2} \epsilon_{x}^{2-d}.
\eeq
\end{itemize}

In both limiting cases the 3-point function diverges for $d>2$ as $\bar{x}_1\to \bar{x}_2$ and the limits of the scaling
function $\Phi(y)$ determine the concrete singularity behavior.

The last expression, \eqref{pm8}, is actually in agreement with the usual requirement for a wave function to describe
fermions at unitarity \cite{NishidaSon}: the wave-function of $N$ spin-up and $M$ spin-down fermions $\Psi(\vec{x}_1,
\dots, \vec{x}_N; \vec{y}_1, \dots, \vec{y}_M)$ behaves like $|\vec{x}_i - \vec{y}_j|^{2-d}$ when $|\vec{x}_i -
\vec{y}_j| \rightarrow 0$ for any pair of fermions with opposite spins $i$, $j$. This simply follows from the scaling
dimension of the operators at unitarity. So that we can take these results as reassuring of being at the unitary regime
with which the holographic computation agrees.

\section{Conclusions}
\label{sect:concl}

In this work we have explored the computation of the n-point scalar correlation functions in the framework of
non-relativistic holography in the vacuum state, i.e.~at zero temperature and density, of the theory defined as the
holographic dual of the Schr\"odinger metric with $z=2$ \cite{Son, McGreevy1}. Following the standard holographic
dictionary the correlators can be expressed in terms of Witten diagrams. At tree level we have demonstrated how the
computation in Sch$_{d+3}$ can be done equivalently from the correlation functions of pure AdS after a projection of the
non-relativistic mass momentum in every insertion at the boundary in the light-cone frame.
\begin{equation}
 \langle \phi_1(\bar{x}_1) \dots \phi_n(\bar{x}_n) \rangle_{\text{Sch}}^{(\text{tree level})} = \int \prod_{k=1}^n d\xi_k\, e^{-i\sum_{j=1}^n ( M_j\xi_j )} \langle
 \phi_1(\bar{x}_1, \xi_1) \dots \phi_n(\bar{x}_n, \xi_n)  \rangle_{\text{AdS}}^{(\text{tree level})}
\end{equation}
This is a useful trick since AdS amplitudes have already been well studied in the literature. The mapping works
irrespectively of whether we have a compact or non-compact $\partial_\xi$-direction. It shows that all observables at
tree level are going to agree in the Sch$_{d+3}$ construction and the pure AdS construction in light-cone coordinates
with definite $\partial_\xi$-momentum, as was loosely noted in \cite{Goldberger, Barbon}. This tree-level mapping can be
understood from the QFT side to be in the same footing as the agreement of the correlation functions at the planar level
between a non-commutative and commutative QFT, since, at least for $d=2$, one can argue \cite{Maldacena} that the
Schr\"odinger background is dual to a non-commutative version of $\mathcal{N}=4$ SYM, the dual of AdS. At the loop
level, or what is the same $1/N$ corrections, we do not have this mapping anymore. In this case we have no option but to
perform all the computations directly in the Schr\"odinger background, which is the only one that has the correct
non-relativistic causal structure \cite{Rangamani}. At the loop level we also see explicitly the need for regularization
of the theory when the $\partial_\xi$-direction is compact. In this case, we have a null-compact direction and the zero
modes along that direction cause the typical divergences one finds in DLCQ theories \cite{Hellerman}. One way to
regularize it is by the introduction of a non-zero chemical potential that makes the circle to be space-like. When the
$\partial_\xi$-direction is non-compact this problem is not present.

We have tested the tree-level mapping to AdS by the computation of the 2-point scalar function finding agreement with
the expected result completely fixed by symmetry considerations, for both compact and non-compact
  $\partial_\xi$-direction.


We have also computed the holographic 3-point scalar function for both compact and non-compact
  $\partial_\xi$-direction. The result respects the form dictated by the Schr\"odinger symmetry although it is not
completely fixed by it. There is freedom for an unknown scaling function.  Remarkably, the form of this function
coincides with the result coming from the theory of cold atom at unitarity, that we have also computed. Upon closer
examination, this function governs the singular behaviour when two operators approach. And it reproduces, since it is
the same as the theory of cold atoms at unitarity, the expected singular behaviour in the unitary regime.  We see this
as a non-trivial check that the holographic theory really contains a conformal non-relativistic theory in the unitary
regime.  

Our computations at tree level worked well in both cases where $\xi$ is periodic or not. This is understandable: the
compactification procedure does not break any spacetime symmetry and the Bargmann superselection rule, which is
characteristic for the non-relativistic systems, is valid in both cases. On the one hand, the compactification of $\xi$
leads naturally to the discreteness of the mass spectrum of a simple one-species system. On the other hand, working with
the non-compact $\xi$ allows to describe the systems with more than one species of particles, that is evidently of great
interest in the cold atoms physics. This last scenario should not be rejected. At tree level with a non-compact $\xi$,
one can consistently restrict by hand the different values of the masses out of the possible continuum and obtain
physical sound answers by virtue of the Bargmann superselection rule.

 \smallskip

Our work here has just scratched the surface of many more interesting questions awaiting to be addressed. Among them we
can consider:

\begin{itemize}

\item It would be interesting to study the higher-point (especially 4-point) correlation functions in the framework of
  the non-relativistic AdS/CFT. Although the functional constraints, implied by the Schr\"odinger symmetry, are not
  known for the higher-point function so far, it seems straightforward to apply our prescription to the 4-point scalar
  correlator already known in AdS.

\item Recently the non-relativistic AdS/CFT was extended to fermionic fields \cite{Akhavan}. Since the original
  motivation of the non-relativistic holography were two-component fermions at unitarity, it is tempting to study
  general n-point functions for holographic fermions. Our expectations are that one should be able to demonstrate the
  mapping of the computations of the fermionic correlation functions in Sch to AdS at tree level. Actually the case of
  2-point functions was explicitly shown to work this way in \cite{Akhavan}.

\item Another interesting question would be the computation of the correlation functions at finite chemical potential in the
  background \eqref{eq:metricT0mu}. 

\item One should also try to address the working of the regularization at the loop-level in the compact
    $\xi$ case.

\end{itemize}

We leave these questions for future study.

\subsection*{Acknowledgements}
\label{acks}

We would like to thank Jos\'e L.~F.~Barb\'on, Peter Horvathy, Djordje Minic, Michel Pleimling, Mukund Rangamani and
Mirko Rokyta for encouragement and useful discussion. We also thank the referee for useful comments.
C.A.F.~is supported by a FPU fellowship from MEC under grant AP2005-0134. S.~M.~is supported by Klaus Tschira
scholarship from KTF.

\appendix

\section{Projected bulk-to-boundary propagator in position space}
\label{app:A}
In the relativistic AdS/CFT the bulk-to-boundary propagator in position space can be computed employing the discrete
inversion symmetry of AdS space \cite{AdSCFT}. The Schr\"odinger spacetime (\ref{eq:sch_metric1}) does not posses this
symmetry. Nonetheless, we can compute the projected propagator as introduced in Sec.~\ref{sect:n-point},
Eq.~\eqref{prbtb}. Here we will present a derivation of the explicit expression for it,
\begin{equation} \label{eq:propagatorApp}
 K_M (\bar{x}- \bar{y},\xi - \xi', u_x) = {e^{iM (\xi - \xi')}} \int d\tilde{\xi}\,
 K_\Delta^{(\text{AdS})}(\bar{x}-\bar{y}, \tilde{\xi}, u_x) \, e^{-iM\tilde{\xi}} \quad.
\end{equation}

In order to get the non-relativistic bulk-to-boundary propagator we must Fourier transform the well-known AdS
bulk-to-boundary propagator \cite{AdSCFT} to a fixed $\xi$-momentum, $M$, in the Euclidean light-cone frame
\beq \label{btb11} \mathscr{K}_{M}(u,\bar{x})=C_{\Delta}\int_{-\infty}^{\infty}d\xi \exp[-iM\xi]
\left(\frac{u}{u^2+\vec{x}^2-2it\xi}\right)^{\Delta} \quad, \eeq with $C_{\Delta} =
\frac{\Gamma(\Delta)}{\pi^{\frac{d+2}{2}}\Gamma[\Delta-\frac{d+2}{2}]}$ and $\Delta(\Delta - d - 2) = m_0^2 + \beta^2 M^2$.

 The integral in Eq.~(\ref{btb11}) was done in Appendix B
\cite{Henkel2} and we follow these calculations closely
\begin{eqnarray} \label{btb12}
\mathscr{K}_{M}(u,\bar{x})&=&C_{\Delta}\left(\frac{u}{2it}\right)^{\Delta}\int_{-\infty}^{\infty}d\xi \exp[-iM\xi]\frac{1}{(\xi+\frac{i(u^2+\vec{x}^2)}{2t})^{\Delta}}= \nonumber \\
&&=C_{\Delta}\left(\frac{u}{2it}\right)^{\Delta} (M)^{\Delta-1} \int_{-\infty}^{\infty}d\xi \exp[-i\xi]\frac{1}{(\xi+\frac{iM(u^2+\vec{x}^2)}{2t})^{\Delta}} = \nonumber \\
&&=C_{\Delta}\left(\frac{u}{2it}\right)^{\Delta} (M)^{\Delta-1} \exp[-\frac{M}{2}\frac{u^2+\vec{x}^2}{t}]\underbrace{\int_{\mathbf{R}+\frac{iM}{2}\frac{u^2+\vec{x}^2}{t}}d \xi e^{-i\xi}\xi^{-\Delta}}_{\alpha \theta(t)}= \nonumber \\
&&=\gamma \left(\frac{u}{t} \right)^{\Delta} \theta(t) \exp[-\frac{M}{2} \frac{u^2+\vec{x}^2}{t}],
\end{eqnarray}
where we introduced a constant $\gamma=C_{\Delta}\alpha (2i)^{\Delta}M^{\Delta-1}$. It is simple to check that Eq.~(\ref{btb12})\footnote{without the causal factor $\theta(t)$} indeed solves the scalar field equation (\ref{eq:op_proj_prop_Sch}) provided that the condition (\ref{ident}) holds.

It is instructive to investigate the behavior of the propagator near the ``boundary'' ($u\to 0$). The function $\mathscr{K}_{M}(u,\bar{x})$ has two important properties
\begin{itemize}
\item $\mathscr{K}_{M}(u,\bar{x})$ has its support at the origin of the ``boundary space'', i.e. at $t=0$ and $\vec{x}=0$, in the limit $u\to 0$.
\item The integral of $\mathscr{K}_{M}(u,\bar{x})$ over the boundary coordinates $t$ and $\vec{x}$ has the form
\beq \label{btb14}
\int d \bar{x} \mathscr{K}_{M}(u,\bar{x})= u^{d+2-\Delta} I,
\eeq
where $I$ is independent of the radial coordinate $u$
\beq
I=\gamma \int d \bar{x} \exp[-\frac{M}{2}\frac{1+\vec{x}^2}{t}].
\eeq
\end{itemize}
These two properties imply that near the ``boundary'' ($u\to 0$) the function
\beq \label{btb14a}
\mathscr{K}_{M}(u,\bar{x})\to u^{d+2-\Delta}\delta(\vec{x})\delta(t)
\eeq
which is a correct behavior of the bulk-to-boundary propagator in the position space. 

The explicit expression of the projected bulk-to-boundary propagator is then \beq \label{btb17}
\begin{split}
 K_{M}(\bar{x}-\bar{y}, \xi-\xi^{\prime},u) & = e^{iM(\xi-\xi^{\prime})}\mathscr{K}_{M}(u, \bar{x}-\bar{y})= \\
& = \gamma e^{iM(\xi-\xi^{\prime})} \left(\frac{u}{\Delta t} \right)^{\Delta} \theta(\Delta t) \exp[-\frac{M}{2} \frac{u^2+\Delta\vec{x}^2}{\Delta t}]\\
\end{split}
\eeq
where $\Delta t= t_{x}-t_{y}$ and $\Delta \vec{x}=\vec{x}-\vec{y}$.

\section{Inverse Fourier transformation} \label{Fourier}
In this Appendix we present a calculation of the inverse Fourier transformation of a general non-relativistic scale-invariant propagator in $d$ spatial dimensions
\beq \label{F1}
G(t, \vec{x})=\int \frac{d \omega}{2\pi} \frac{d^{d}q}{(2\pi)^d} (i \omega+\epsilon_{\vec{q}})^{\alpha} e^{i(\omega t - \vec{q} \cdot \vec{x})},
\eeq
where $\epsilon_{\vec{q}}=\frac{\vec{q}^2}{2M}$ with the non-relativistic mass $M$. We assume that $\alpha$ is real and negative.

First, the $d-1$ dimensional angular integration can be performed employing a general formula
\beq \label{F2}
\int \frac{d^{d}q}{(2\pi)^d} f(q) e^{-i\vec{q}\cdot \vec{x}}=\left(\frac{1}{2\pi} \right)^{\frac{d}{2}} \int_{0}^{\infty} dq f(q) q \left(\frac{q}{x}\right)^{\frac{d}{2}-1}J_{\frac{d}{2}-1}(qx),
\eeq
which is valid for an arbitrary radial function $f(q)$. In our case $f(q)=(i \omega+\epsilon_{\vec{q}})^{\alpha}$ and we arrive at
\beq \label{F3}
G(t, x)=\left(\frac{1}{2\pi} \right)^{\frac{d}{2}} \int_{0}^{\infty} dq q \left(\frac{q}{x}\right)^{\frac{d}{2}-1}J_{\frac{d}{2}-1}(qx) \int \frac{d\omega}{2\pi} e^{i\omega t} (i\omega +\epsilon_{\vec{q}})^{\alpha},
\eeq
where $x^2=\vec{x}\cdot \vec{x}$. 

The frequency integration in Eq.~(\ref{F3}) can be done by performing a substitution $i\omega\to i \omega +\epsilon_{\vec{q}}$
\beq \label{F4}
G(t, x)=(i)^{\alpha}\left(\frac{1}{2\pi} \right)^{\frac{d}{2}} \int_{0}^{\infty} dq q \left(\frac{q}{x}\right)^{\frac{d}{2}-1}J_{\frac{d}{2}-1}(qx) e^{-\frac{q^2 t}{2M}} \int_{\mathds{R}-i\frac{q^2}{2M}} \frac{d\omega}{2\pi} e^{i\omega t} \omega^{\alpha}.
\eeq
The integrand has a branch point at $\omega=0$. The branch cut goes from zero to infinity and we choose it to be along
the positive imaginary axis. This choice corresponds to the retarded causal structure of the propagator. Due to this
singularity structure of the integrand, the integration contour in Eq.~(\ref{F4})  can be shifted to the real axis. 
\begin{figure}[t]
\begin{center}
\includegraphics[height=40mm]{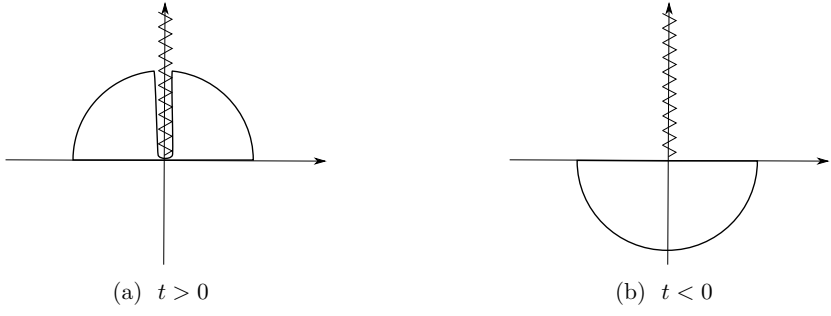}
\end{center}
\caption{Contour of integration for $t>0$ and $t<0$. The branch cut of $\omega^{\alpha}$ was chosen to be along the positive imaginary axis.}
\label{fig3}
\end{figure}


If $t>0$, we can consider a contour like the one in Fig.~\ref{fig3} (a). Applying the residue theorem we obtain that the integral along the real axis is related to an integral along the positive imaginary axis by
\begin{equation} \label{F5}
\begin{split}
\int_{-\infty}^{\infty}\frac{d\omega}{2\pi} \omega^{\alpha} e^{i\omega t} & = -i^{\alpha+1}\int_{0}^{\infty}\frac{d\chi}{2\pi} \chi^{\alpha} e^{-\chi t}-i^{\alpha+1}e^{2\pi i \alpha}\int_{\infty}^{0}\frac{d\chi}{2\pi} \chi^{\alpha} e^{-\chi t}  \\
&=i^{\alpha+1}(e^{2\pi i \alpha}-1)\int_{0}^{\infty}\frac{d\chi}{2\pi} \chi^{\alpha} e^{-\chi t},
\end{split}
\end{equation}
where we took $\chi=i\omega$. Now applying that
\beq \label{F6}
\int_{0}^{\infty}\frac{d\chi}{2\pi} \chi^{\alpha} e^{-\chi t}=\frac{\Gamma(\alpha+1)t^{-\alpha-1}}{2\pi}=-\frac{t^{-\alpha-1}}{2\Gamma(-\alpha)\sin(\pi \alpha)}
\eeq
we obtain
\beq \label{F7}
\int_{-\infty}^{\infty}\frac{d\omega}{2\pi}\omega^{\alpha}e^{i\omega t}=i^{\alpha} e^{i\pi \alpha}\frac{t^{-\alpha-1}}{\Gamma(-\alpha)}.
\eeq
If $t<0$, we have to consider the contour in Fig.~\ref{fig3} (b). It is clear that the integral is then zero. Hence
\beq \label{F8}
\int_{-\infty}^{\infty}\frac{d\omega}{2\pi}\omega^{\alpha}e^{i\omega t}=i^{\alpha} e^{i\pi \alpha}\frac{t^{-\alpha-1}}{\Gamma(-\alpha)}\theta(t).
\eeq
Putting this into Eq.~(\ref{F4}) we obtain
\beq \label{F9}
G(t, x)=\frac{1}{\Gamma(-\alpha)( 2\pi)^{\frac{d}{2}}} t^{-\alpha-1}\theta(t) \underbrace{\int_{0}^{\infty} dq q \left(\frac{q}{x}\right)^{\frac{d}{2}-1}J_{\frac{d}{2}-1}(qx)e^{-\frac{q^2 t}{2M}}}_{F(t,x)}.  
\eeq
Under the assumptions $t>0$ and $x>0$, the momentum integral in Eq.~(\ref{F9}) can be done analytically
\beq \label{F10}
F(t,x)=\left(\frac{t}{M} \right)^{-\frac{d}{2}} e^{-\frac{Mx^2}{2t}}.
\eeq
Hence, we obtain the final result
\beq \label{F11}
G(t, x)=\left( \frac{M}{2\pi} \right)^{\frac{d}{2}}\frac{\theta(t)}{\Gamma(-\alpha)}t^{-\Delta}e^{-\frac{Mx^2}{2t}},
\eeq
where we defined the scaling dimension $\Delta=1+\frac{d}{2}+\alpha$.

\section{2 and 3-point functions for compact $\partial_\xi$} \label{app:compact}

In this Appendix we will give the computation of the 2 and 3-point function for a compact $\partial_\xi$-direction using
the general relation \eqref{eq:general_rel} discussed in the main text. According to \eqref{eq:general_rel} we have to
perform the Fourier transform of the corresponding (relativistic) point function. For the computation to be consistent
we need to start from the relativistic n-point function that is fully periodic in $\xi_k$.

\subsection{2-point function}

Let us consider the 2-point function when one takes $\xi$ to be compact, $\xi \in [0, L]$. One needs to have a periodic
relativistic 2-point function in $\xi$. The usual one is clearly not
\begin{equation}
  \langle \phi_1(0) \phi_2(\bar{x}, \xi) \rangle = \frac{C_{12}}{ (-2i \xi t + x^2)^\Delta} \;.
\end{equation} 
But one can construct easily one that is indeed periodic in $\xi$, just take
\begin{equation}
  \langle \phi_1(0) \phi_2(\bar{x}, \xi) \rangle = \sum_{n\in \mathds{Z}} \frac{C_{12}}{ (-2i t ( \xi + nL) +
    x^2)^\Delta} \;.
\end{equation}
This is a simple linear combination of 2-point functions and, hence, it is going to fulfill the differential equations
for the propagator that are linear.

This way the 2-point function when we have the periodically identified $\xi$ is (this is the exact equivalent of
equation (52))
\begin{equation}
  \begin{split}
    G_2 & = L \delta_{M_1 + M_2, 0} \int_{-\frac{L}{2}}^{\frac{L}{2}} \langle \phi_1(X_1) \phi_2(X_2) \rangle_{\text{AdS}}^{\text{(tree level)}} \\
        & = \frac{L C_{12} \delta_{\Delta_1, \Delta_2} \delta_{M_1 + M_2, 0}}{ 2i(t_1 - t_2)^{\Delta_1}} \int_{-\frac{L}{2}}^\frac{L}{2} d\xi
        e^{-iM_1 \xi} \sum_{n\in \mathds{Z}} \left( \xi + nL + \frac{i(\vec{x}_1 - \vec{x}_2)^2}{2(t_1 - t_2)} \right)^{-\Delta_1}
  \end{split}
\end{equation}
Here the masses are quantized as $M_i = 2\pi j/L$ with $j \in \mathds{Z}$.

Now if we do the simple change of variables
\begin{equation}
\tilde{\xi} = \xi +  \frac{i(\vec{x}_1 - \vec{x}_2)^2}{2(t_1 - t_2)}  
\end{equation}
one finds
\begin{equation}
  G_2 =  \frac{L C_{12} \delta_{\Delta_1, \Delta_2} \delta_{M_1 + M_2, 0}}{ 2i(t_1 - t_2)^{\Delta_1}}
    e^{ \frac{-M_1 (\vec{x}_1 - \vec{x}_2)^2}{2(t_1 - t_2)}  }\int_{ {-\frac{L}{2}} + \frac{i(\vec{x}_1 - \vec{x}_2)^2}{2(t_1 - t_2)}
    }^{{\frac{L}{2}} +  \frac{i(\vec{x}_1 - \vec{x}_2)^2}{2(t_1 - t_2)}  } d\tilde{\xi}
        e^{-iM_1 \tilde{\xi}} \sum_{n \in \mathds{Z}} \left( \tilde{\xi} + nL \right)^{-\Delta_1}
\end{equation}
The last integral is proportional to $\theta(t_1 - t_2)$ and otherwise independent of $\vec{x}$ and $t$. Hence we see that the
final result agrees with Schr\"odinger symmetry, Eq.~(\ref{u3}).

\subsection{3-point function}

The 3-point function, $G_3$, in a (relativistic) conformal field theory is completely fixed (up to an overall constant)
by the symmetries of the theory. That is, it is completely fixed by a set of linear partial differential equations that
come from imposing invariance under the generators of the conformal group. As such, applying an identical procedure as
in the 2-point function, in order to find the 3-point function that is periodic in the different $\xi_j$ all we have to
do is to substitute $G_3(\xi_j)$ by $\sum_{n_j} G_3(\xi_j + Ln_j)$.


Starting then from \eqref{3ph1} and following the same manipulations as in the non-compact $\xi$ case, it is immediate
that the spacetime dependence will be exactly the same as in the non-periodic $\xi$ case. On the other hand the
expression for the scaling function that one has to evaluate, reduces to
\begin{equation}
  \Phi(y) = \int_{{-\frac{L}{2}}+i\varepsilon}^{{\frac{L}{2}} + i\varepsilon} dv \int_{{-\frac{L}{2}}+i\varepsilon'}^{{\frac{L}{2}} + i\varepsilon'} dv' e^{-iM_1 v - iM_2 v'}
  \sum_{n,n'\in \mathds{Z}} (v - v' + Ln - Ln' + iy)^{-\frac{\Delta_{12,3}}{2}} (v + Ln)^{-\frac{\Delta_{13,2}}{2}}  (v' + Ln')^{-\frac{\Delta_{23,1}}{2}}
\end{equation}
with $\varepsilon, \varepsilon'>0$.

For the case, $\Delta_1 = \Delta_2 = d/2$, $\Delta_3 = 2$, $M_1 = M_2 = 1$, the integral becomes
\begin{equation} \label{eq:peridic_psi}
  \Phi(y) = \int_{{-\frac{L}{2}}+i\varepsilon}^{{\frac{L}{2}} + i\varepsilon} dv \int_{ {-\frac{L}{2}} +i\varepsilon'}^{
    {\frac{L}{2}}+ i\varepsilon'} dv' e^{-iv - i v'}
  \sum_{n,n'\in \mathds{Z}} (v - v' + Ln - Ln' + iy)^{-\frac{d-2}{2}} (v + Ln)^{-1}  (v' + Ln')^{-1}
\end{equation}

In order to evaluate \eqref{eq:peridic_psi}, we apply the residue theorem together with a contour of integration that
consists on a rectangle with base $[-L/2,L/2]$ and height extending towards minus the imaginary infinity for both
integrals. We first do the integral in $v'$. Assuming $y>0$, there is only one pole that lies within the contour of
integration, the one at $v' = 0$.  The residue at $v'=0$ is
\begin{equation}
 \begin{split}
  \lim_{v' \rightarrow 0} 
  \sum_{n,n'\in \mathds{Z}} \frac{v'   e^{-iv - i v'}}{(v - v' + Ln - Ln' + iy)^{\frac{d-2}{2}} (v + Ln)  (v' + Ln')} 
    = \sum_{n\in \mathds{Z}} \frac{ e^{-iv}}{(v + Ln  + iy)^{\frac{d-2}{2}} (v + Ln)}
\end{split}
\end{equation}
Thus
\begin{equation}
    \Phi(y) = - 2\pi i \int_{{-\frac{L}{2}} +i\varepsilon}^{{\frac{L}{2}} + i\varepsilon}  \sum_{n\in \mathds{Z}} \frac{ e^{-iv}}{(v + Ln  +  iy)^{\frac{d-2}{2}} (v + Ln)} \;.
\end{equation}
Applying again the residue theorem with the same contour of integration and the relations \eqref{comp2} or \eqref{eq:int_rep_gamma}
depending on whether is $d$ is even or odd respectively, one arrives to
 \begin{equation}
\begin{split}
    \Phi(y) = N_d\, y^{-\frac{d}{2}+1} \gamma(\frac{d}{2} - 1, y) \;.
\end{split}
\end{equation}

We see that the only difference with respect to the non-compact $\xi$ case is that $L$ appears as a multiplicative
factor of the 3-point function instead of $2\pi$. The scaling function is exactly the same as in the non-compact case,
Eq.\eqref{comp3},

\end{document}